\documentclass[pra,aps,twocolumn,eqsecnum,superscriptaddress,showpacs]{revtex4-1}

\usepackage{amsfonts}
\usepackage{amsmath, amssymb}
\usepackage{bm} 
\usepackage{graphicx} 
\usepackage{epsfig}
\usepackage{epstopdf}
\usepackage{bm} 
\usepackage{color}
\usepackage{float}
\usepackage{ulem}
\usepackage{hyperref} 
\hypersetup{colorlinks=true,citecolor={blue},linkcolor={blue},urlcolor={blue},pdfstartview=FitH}
\usepackage{ulem}
\usepackage[colorinlistoftodos]{todonotes}

\newcommand{\veps}{\varepsilon}

\renewcommand{\Im}{\mbox{Im}}

\graphicspath{{pics/}}

\begin{document}

\title{Extremely subradiant states in a periodic one-dimensional atomic array}

\author{D. F. Kornovan}
\email{d.kornovan@metalab.ifmo.ru}
\affiliation{ITMO University, Birzhevaya liniya 14, 199034 St.-Petersburg, Russia}
\author{N. V. Corzo}%
\affiliation{Laboratoire Kastler Brossel, Sorbonne Universit\'e, CNRS, ENS-Universit\'e PSL, Coll\`ege de France, 4 place Jussieu, 75005 Paris, France}
\affiliation{Centro de Investigaci\'{o}n y de Estudios Avanzados del I.P.N. Unidad Queretaro, Libramiento Norponiente 2000, 76230 Queretaro, Mexico}
\author{J. Laurat}%
\affiliation{Laboratoire Kastler Brossel, Sorbonne Universit\'e, CNRS, ENS-Universit\'e PSL, Coll\`ege de France, 4 place Jussieu, 75005 Paris, France}
\author{A. S. Sheremet}%
\email{alexandra.sheremet@lkb.upmc.fr}
\affiliation{Laboratoire Kastler Brossel, Sorbonne Universit\'e, CNRS, ENS-Universit\'e PSL, Coll\`ege de France, 4 place Jussieu, 75005 Paris, France}
\affiliation{Russian Quantum Center, Novaya 100, 143025 Skolkovo, Moscow Region, Russia}

\date{\today}

\begin{abstract}

We study the subradiant collective states of a periodic chain of two-level atoms with either transversal or longitudinal transition dipole moments with respect to the chain axis. We show that long-lived subradiant states can be obtained for the transversal polarization by properly choosing the chain period for a given number of atoms in the case of no open diffraction channels. While not being robust against the positional disorder along the chain, these highly subradiant states have a linewidth that decreases with the number of atoms much faster than it was shown previously. In addition, our study shows that similar states are present even if there are additional interaction channels between the atoms, i.e. they interact via waveguide mode of an optical nanofiber for instance. We develop a theoretical framework allowing to describe the spectral properties of the system in terms of contributions from each collective eigenstate and we show that subradiant states manifest themselves in the transmission and reflection spectra, allowing to observe interaction-induced transparency in a very narrow spectral range. Such long-lived collective states may find potential applications in nanophotonics and quantum optics. 

\end{abstract}

\maketitle

\section{Introduction}

Cooperative effects in spatially dense atomic ensembles have generated a large interest in recent years due to the significant induced modifications to the optical properties of the system \cite{Dicke1954, Lehmberg1970, Andreev1980, Gross1982, Sokolov2009, Ritsch2015, Robicheaux2016, Guerin2016, Kupriyanov2017, Kuraptsev2014, Kimble2018}. These effects come from strong dipole-dipole interaction in a collection of quantum emitters with a sub-wavelength average separation. Recent experimental advances in trapping techniques have made it possible to create $1$D \cite{Miroshnychenko2006, Endres2016}, $2$D \cite{Nogrette2014, Barredo2016, Kim2016, deMello2019}, and $3$D \cite{Nelson2007, Barredo2018} spatially-ordered atomic configurations where such collective effects can play a very important role. The most prominent phenomenon is superradiance \cite{Felinto2014, Goban2015, Bromley2016, Araujo2016, Roof2016}, i.e. the enhancement of the collective spontaneous emission rate that can be explained as a constructive interference between the emission pathways of $N$ closely-located dipoles. Contrary to this effect, the subradiance \cite{Kaiser2012, Ritsch2015, Scully2015, GuerinPRL2016, JenPRA2016, ZhangPRL2019, ZhangArxiv2019} is the suppression of the collective emission rate due to the destructive interference between dipoles. Being interesting due to their enhanced lifetimes, these states are, however, hard to observe experimentally because of their weak coupling to the light field and strong sensitivity to additional nonradiative decay channels. Nevertheless, such states have been observed for a pair of trapped ions \cite{DeVoe1996}, ultracold molecules \cite{Zelevinsky2015} polymer nanostructures \cite{Barnes2005}, atomic gases \cite{GuerinPRL2016, Weiss2018}, and thermal light sources \cite{Bhatti2018}.

At the same time, one-dimensional systems recently gained a special attention as a possible platform for quantum light-matter interfaces due to the strong transverse confinement of the light field and the possibility of infinite-range atom-atom interaction \cite{Chang2014, Pichler2015}. Such system is a versatile platform for achieving efficient light-atom coupling due to the collective nature of atomic interaction with the evanescent field of the guided mode \cite{Balykin2004}. The strong coupling of an atomic ensemble with such a nanophotonic waveguide provides opportunities to further develop the emerging field of waveguide-QED \cite{ZhengPRL2013, Shen2007, GuPhysRep2017, Kimble2018}, in which many remarkable results were recently demonstrated not only in the field of theoretical research \cite{Qi2016, Liao2017, Liao2015, Roy2017, Mahmoodian2018, Shen2007, Gonzales-Tudela2017, Paulisch2016, KornovanPRB2016, Kornovan2017, ChangPRX2017, VahidPRA2016, Pivovarov2018} but also in experiments \cite{Vetsch2010, Goban2012, Gouraud2015, Corzo2016, Sorensen2016, Beguin2018, CorzoNature2018}, including observation of subradiant states \cite{Solano2017}.

From these perspectives subradiant states in quasi one-dimensional atomic chains are of interest for the development of new approaches in quantum technologies. In particular, generation of a periodic one-dimensional atomic chain in the sub-diffractional regime, where the period of the system is smaller than half of the resonant wavelength, can bring new effects that cannot be achieved in free space because of the limited mode matching between light and the atomic system \cite{VahidPRA2016, ChangPRX2017}.

Optical properties of $1$D systems containing a large number of scatterers were studied previously in different contexts starting from arrays of nanoparticles  \cite{WeberFord2004, RasskazovJOSAB14, MarkerJoMO1993, MarkelJOPB2005} to cold atoms \cite{BettlesPRA2016, SutherlandPRA2016}. In this work, we study the subradiant effects occurring in a periodic one-dimensional atomic chain in the subwavelength regime, when the period of the system is smaller than $\lambda_0/2$, with $\lambda_0 = 2\pi c/\omega_0$ being the resonant radiation wavelength. We consider a regular $1$D chain of two-level atoms coupled to a single-mode nanofiber including free-space spontaneous emission with inherent dipole-dipole coupling, as shown in Fig. \ref{fiber}. 
In order to study the effect of subradiance, we have extended the formalism developed in our previous work \cite{KornovanPRB2016} introducing the eigenstate picture \cite{MarkelJOSAB1995}, and calculated transmission and reflection coefficients for each eigenstate. In our theoretical formalism, we consider the full Green’s tensor of the electromagnetic field with taking into account all of the field modes (free-space, radiation, guided, and near-field modes) without applying the paraxial regime as it was done in \cite{Pivovarov2018}. As a reference point, we first study suppression of spontaneous emission rate for the atomic chain in the vacuum using a microscopic approach of light scattering. In a further step, we extend our system considering atoms trapped near a single-mode nanofiber. In such a system in addition to the collective coupling to the 3D free-space vacuum modes, the nanofiber structure introduces an additional channel of virtually infinite-range dipole-dipole coupling. In this work we aim to find conditions required for manifestation of a highly subradiant state with the collective emission rate less than the known $N^{-3}$ scaling.

The paper is organized as follows. In Sec. II we first remind the theoretical framework for the case of a periodic chain in free-space and discuss different quantities relative to the studied collective effect. For the sake of intuitive understanding, we start consideration for $N = 3$ atoms and then increase number of atoms in the system. In Sec. III we discuss the modification of the theory for the case of the waveguide mode scattering and demonstrate a similar expansion to the one developed in Sec. II. We show that the long-lived dark states are present even for atoms coupled not only by a vacuum dipole-dipole interaction but also through a guided mode.

\section{Light scattering on an atomic array in the vacuum}

\begin{figure}[t]
\includegraphics[width=0.48\textwidth]{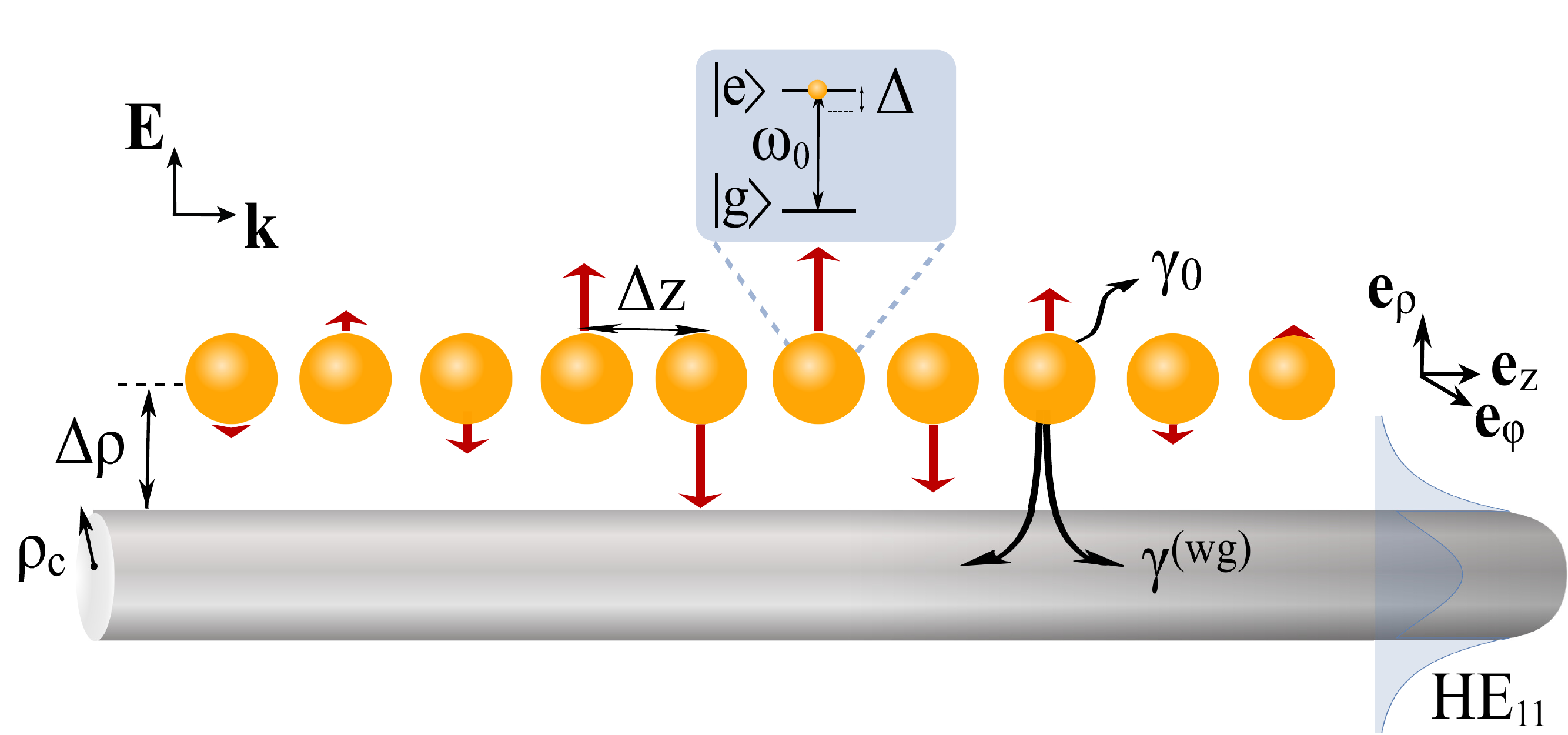}
\caption{Light scattering on the $1$D array of two-level atoms separated by a distance $\Delta z$ and trapped at a distance $\Delta\rho$ from the surface of an optical nanofiber with radius $\rho_c$ and permittivity $\varepsilon$. The fiber radius $\rho_c$ is less than the atomic resonant wavelength, so only the fundamental mode $\text{HE}_{11}$ can be guided. The red arrows indicate the eigenvector components of the subradiant state for $N = 10$ atoms.}
\label{fiber}
\end{figure}

In this section, we consider single-photon scattering in a one-dimensional array of $N$ two-level atoms with a period $\Delta z$ in vacuum, see Fig. \ref{fiber}. A single photon with a near-resonant atomic frequency induces electric dipoles in each atom of the array. The strength of the atomic response on the incident photon drastically depends on the interatomic distance. Thus, atoms with a large distance between its neighbours $\Delta z$ behave as independent scatters, while closely located atoms bring a collective response. The key point of this collective behavior is that each atom is driven not only by the incident photon, but also by the field emitted by all other atoms in the array. The resulting dipole-dipole interaction between atoms significantly modify their scattering properties. 

In quantum theory, the scattering process can be described in terms of the scattering matrix which can be linked to observable variables such as transmittance and reflectance. Moreover, the cooperative nature of the interaction can be roughly characterized by the resonance widths of the total cross-section spectra or the decay rates of the collective states.

In this section, we investigate collective effects by studying the eigenvalues of the system. These characteristics allow us to find decay rates for each collective state and cross section spectrum. The emergence of the collective effects strongly depends on the interatomic distance $\Delta z$, thus illustrating the role of dipole-dipole interaction in the formation of subradiant states.

\subsection{General theoretical formalism}

In order to simplify the theoretical description, let us consider single-photon scattering process, where the initial and the final states of the decoupled atom-photon system can be represented as $| l \rangle = | g \rangle^{\otimes N} |1_{\mu}\rangle$, $|k\rangle = |g\rangle^{\otimes N} |1_{\mu'}\rangle$, where the index $\mu$ describes a particular field mode $\mu = (\mathbf{k},s)$, where $\mathbf{k}$ is the wave vector, $s = 1,2$ denotes two orthogonal polarizations and $|g\rangle^{\otimes N}$ means that all $N$ atoms are in the ground state $|g\rangle$.

The scattering process can be described by the scattering matrix $S$ \cite{ClaudeCohen-Tannoudji2004} that transforms the asymptotic states from the initial $l$ to the final system state $k$ and has the following form:
\begin{eqnarray}
S_{kl} = \delta_{kl} - 2\pi i T_{kl}(E_l+i0)\delta(E_{k}-E_{l}).
\label{Smatgen}
\end{eqnarray}
Here the $T$ matrix has the standard form \cite{ClaudeCohen-Tannoudji2004}:
\begin{equation}
\hat T = \hat V + \hat V \hat G(E+i0) \hat V,
\label{Tmatgen}
\end{equation} 
where $\hat G(E) = \left(E - \hat H\right)^{-1}$ is the resolvent operator of the total Hamiltonian $\hat H = \hat H_0 + \hat V$. In the dipole approximation the interaction operator $\hat{V}$ has the form $\hat V = -\sum_{i=1}^{N}\hat{\mathbf{d}}_i \hat{\mathbf{ E}}(\mathbf{r}_i)$, where $\hat{ \mathbf{d}}_i = d_{i,eg} \hat {\sigma}^{+} + d_{i,ge} \hat{\sigma}^{-}$ is the dipole moment operator of the $i$-th atom, and $\hat{\mathbf{E}}(\mathbf{r}_i)$ is the field operator at the atomic position $\mathbf{r}_i$. In the rotating-wave approximation the matrix elements of the operator $\hat{T}$ can be found as a projection onto the Hilbert subspace of the vacuum state for the electromagnetic field and the single excited state for the atomic subsystem \cite{ClaudeCohen-Tannoudji2004}:
\begin{eqnarray}
\hat P \hat G(E) \hat P &=& \hat P \dfrac{1}{E - \hat H_0 - \hat \Sigma(E)} \hat P,
\nonumber\\
\hat \Sigma(E) &=& \hat V \dfrac{1}{E - \hat H} \hat V,
\label{ProjResGen}
\end{eqnarray}
where the projector operator can be defined as $\hat P = \sum_{i = 1}^{N}|g_1,...,e_i,...g_N;\{0_{\mu}\}\rangle \langle \{0_{\mu}\};g_1,...e_i,...,g_N|$, and the level-shift operator has the form $\hat \Sigma(E) \approx \hat V (E - \hat H_0)^{-1}\hat V$ in second-order perturbation theory. 

We now apply the resonant approximation, where the scattering photon frequency $\omega$ can be considered close to the atomic transition frequency $\omega_0$. In this approximation the level-shift operator $\hat{\Sigma}(E)$ can be assumed to be a slowly varying function of the argument as $\hat{\Sigma} \approx \hat{\Sigma}(E_0 = \hbar\omega_0)$. The single- and double-particle contributions to the level-shift operator can be written as:
\begin{eqnarray}
\Sigma^{(nn)}(E_0) &=& \hbar\left(\Delta_{\text{L}}^{\text{vac}} - i\frac{\gamma_0}{2}\right)
\nonumber\\
\Sigma^{(mn)}(E_0) &=& -\mathbf{d}_{m, eg} \bigg[\frac{k_0 ^2 e^{i k_0 R}}{R}\bigg(\left(1 + \frac{ik_0 R - 1}{k_0^2 R^2}\right)\mathbf{I}
\nonumber\\
&+& \frac{\mathbf{R}\otimes\mathbf{R}}{R^2}\cdot\frac{3 - 3ikR - k^2R^2}{k^2R^2} \bigg) \bigg] \mathbf{d}_{n, ge}, {\:\:\:\:}
\label{sigmavac}
\end{eqnarray}
where $\Delta_{\text{L}}^{\text{vac}}$ is the vacuum Lamb shift, $\gamma_0 = \frac{4|\mathbf{d}|^2\omega_0^3}{3\hbar c^3}$ is the free-space spontaneous emission rate for a two-level atom, $k_0=\omega_0/c$ is the resonant wavenumber, $\mathbf{R} = |\mathbf{r}_i - \mathbf{r}_j|$ is the distance between an atom \textit{i} and an atom \textit{j}, $\mathbf{I}$ is the unit dyad, and $\otimes$ stands for the outer product. Note that Eq. \eqref{sigmavac} is written in CGS units, and it will be used from now on. Here the single-particle contribution is responsible for the vacuum Lamb shift (and it is considered to be already included into the definition of the transition frequency $\omega_0$) and the finite lifetime of the atomic excited state, while the double-particle contribution describes the excitation transfer between atoms and takes into account the dipole-dipole interaction.

In general, the scattering process in free space is characterized by \textit{the total cross section}, which can be found using  the optical theorem \cite{Sheremet2012}:
\begin{eqnarray}
\sigma_\text{tot} = - \dfrac{2 \mathbb{V}}{\hbar c} \Im\: T_{ii}(E_{i}+i0),
\label{sigmatot}
\end{eqnarray}
where $\mathbb{V}$ is the quantization volume. Since the dipole-dipole interaction alters the eigenstates of the system, it is convenient to expand the total cross-section Eq. \eqref{sigmatot} into a sum, where each term will correspond to a particular collective eigenstate of the system. From Eq. \eqref{ProjResGen} one can see that both $\Sigma(\omega_0)$ and $\left(E - H_0 - \Sigma(\omega_0) \right)^{-1}$ share the same set of eigenvectors, while their eigenstates are simply shifted by $E - H_0$. Therefore, we can rewrite the total cross section taking into account the form of the vacuum field operator $\hat E(\mathbf{r}) = \sum_{\mathbf{k}, s} i \sqrt{\frac{2\pi\hbar\omega}{\mathbb{V}}} \left( \hat a_{\mathbf{k},s} \mathbf{e_{\mathbf{k},s}} e^{i \mathbf{kr}} - h.c. \right)$ as:
\begin{equation}
\sigma_\text{tot}(\Delta) = \sum\limits_{j=1}^{N} \sigma_{j}(\Delta) =  -  \dfrac{3 \pi \hbar\gamma_0}{k_0^2} \Im  \sum\limits_{j=1}^{N} \dfrac{f_{j}}{\hbar\Delta - \lambda_j},
\label{sigmatotexp}
\end{equation}
where $\Delta = \omega - \omega_0$ is the detuning, and  
$f_j = 
\left[\begin{pmatrix}
e^{-i\mathbf{k r_{1}}}, ..., e^{-i\mathbf{k r_{N}}} 
\end{pmatrix} 
S_{\{:,j\}} 
\right] \times \left[ [S^{-1}]_{\{j,:\}} 
\begin{pmatrix}
e^{i\mathbf{k r_{1}}}, ..., e^{i\mathbf{k r_{N}}} 
\end{pmatrix}^{T} 
\right]$ 
with $S$ being the transformation matrix to the eigenspace of $\Sigma(\omega_0)$ with corresponding eigenvectors $S_{\{:,j\}}$ as its columns. The parameter $f_j$ corresponds to a complex-valued oscillator strength amplitude associated with a particular collective eigenstate and for a collection of $N$ two-level atoms $\sum\limits_{j=1}^{N} f_j = N$. The physical meaning of the factor $f_j$, as can be seen from the definition above, is that it is related to the overlap between the photon and the $j^{\text{th}}$ eigenstates of the system. Furthermore from a mathematical point of view, the expansion in Eq. \eqref{sigmatotexp} essentially simplifies the process of finding the total cross section. Thus, instead of inversion of a $N \times N$ matrix for each frequency point for Eq. \eqref{sigmatot}, it is enough to diagonalize the problem only once for a given configuration and then to find the spectrum analytically Eq. \eqref{sigmatotexp}. This property is very important for large number of atoms $N$. Note that similar decomposition was introduced in \cite{MarkelJOSAB1995} to expand the scattering cross section for a collection of classical dipoles.

\begin{figure}[t]
\includegraphics[width=0.48\textwidth]{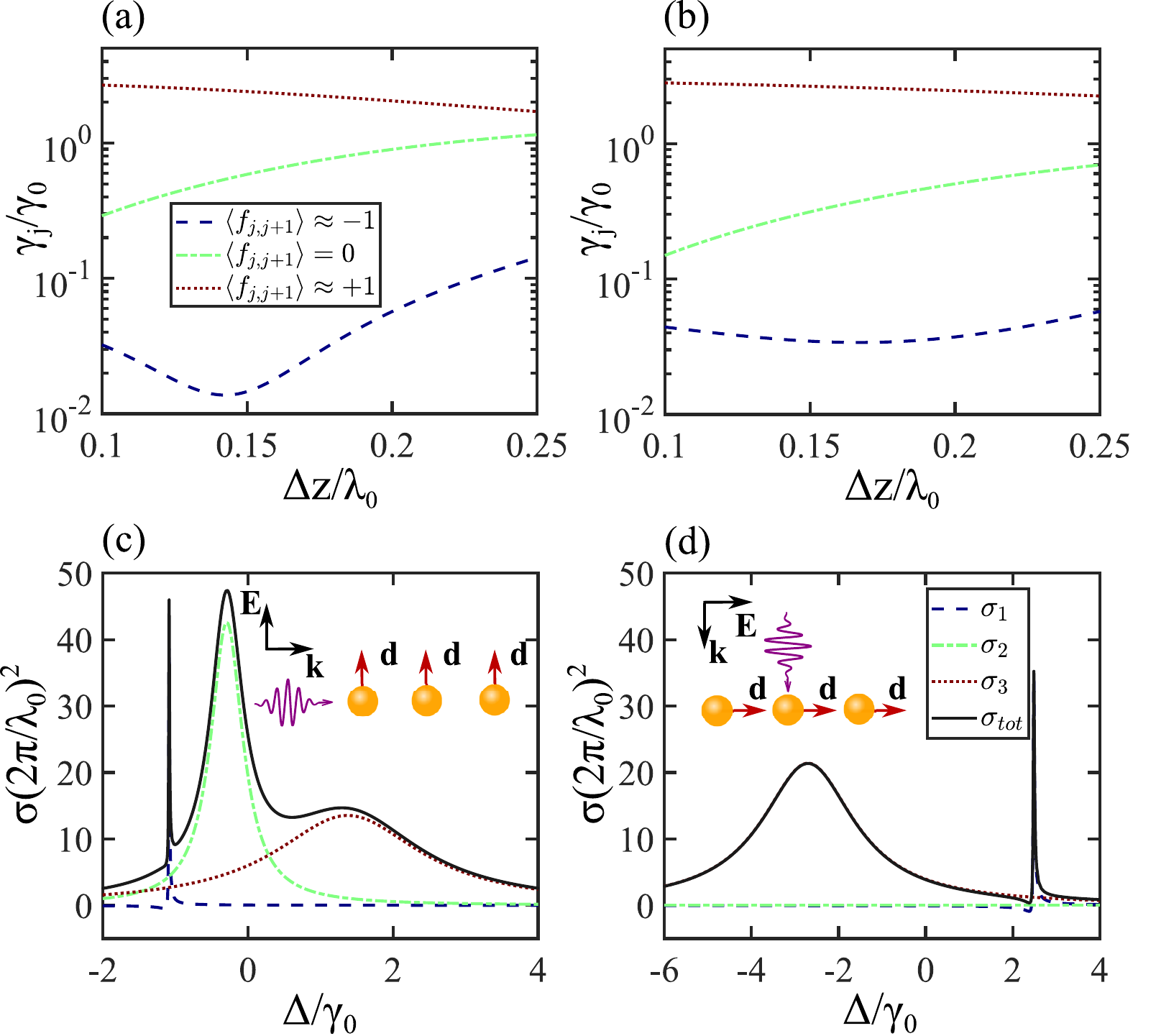}
	\caption{(Color online) Scattering of a photon propagating along (left column) and perpendicular (right column) to the chain axis. (a)-(b) Normalized spontaneous emission rates $\gamma_j$ for the eigenstates of the regular array of $N = 3$ atoms as a function of the period $\Delta z$. Blue dashed, green dashed-dotted, and red dotted lines correspond to three states with different values of the nearest-neighbor correlation function Eq. \eqref{nncf}. (c)-(d) The partial $\sigma_j$ and the total $\sigma_\text{tot}$ scattering cross sections on the array of $N = 3$ atoms with the period $\Delta z$ giving the minimal $\gamma_j$.}
	\label{3atGam}
\end{figure}

We can rewrite the total cross section Eq. \eqref{sigmatotexp} in the following form:
\begin{multline}
\sigma_{\text{tot}}(\Delta) \sim \text{Im} \sum\limits_{j=1}^{N} \left[\dfrac{f_j}{\hbar\Delta - \lambda_j} \right] = \\ \sum\limits_{j=1}^{N} \dfrac{f^{'}_j \lambda'' + f^{''}_j\left( \hbar\Delta - \lambda_j' \right) }{(\hbar\Delta - \lambda_j')^2 + \lambda_j''^2},
\label{ImTExp}
\end{multline}
where  prime and double prime indicate real and imaginary parts respectively. One can see that each contribution $\sigma_j(\Delta)$ to the total cross section consists of two terms: a dissipative term which is proportional to $f'_j$ and has a Lorenzian shape, and a dispersive term, proportional to $f''_j$ that introduces asymmetries. Note that by  analogy with the level-shift operator Eq. \eqref{sigmavac}, here the first term, corresponding to the single particle contribution, is always present, while the second term appears in the system of interacting atoms. As the second term in Eq. \eqref{ImTExp} is antisymmetric, the area under a partial cross section $\sigma_j(\Delta)$ is proportional to $f'_j$: $\int_{-\alpha}^{\alpha} \sigma_{j} (\Delta) d\Delta \sim f'_j$, that provides the contribution of a particular eigenstate to the total cross section. Note that for $\alpha \to \infty$, the corresponding integral $\int\limits_{-\alpha}^{\alpha}\sigma_j(\Delta)d\Delta$ formally diverges, it is a well known problem of the Cauchy distribution having no finite moments of order greater or equal to one. However, we can integrate over a symmetric region with a sufficiently large and physically meaningful parameter $\alpha$. There is also another reason to consider a finite value of $\alpha$: integration over the whole frequency might not be consistent with the Markov approximation ($\hat \Sigma \approx \hat \Sigma (E_0 = \hbar \omega_0)$) in some specific situations.

\subsection{Collective effects in the array of $N = 3$ atoms}

\begin{figure*}[!bt]
\centering
	\includegraphics[width = 0.99\textwidth]{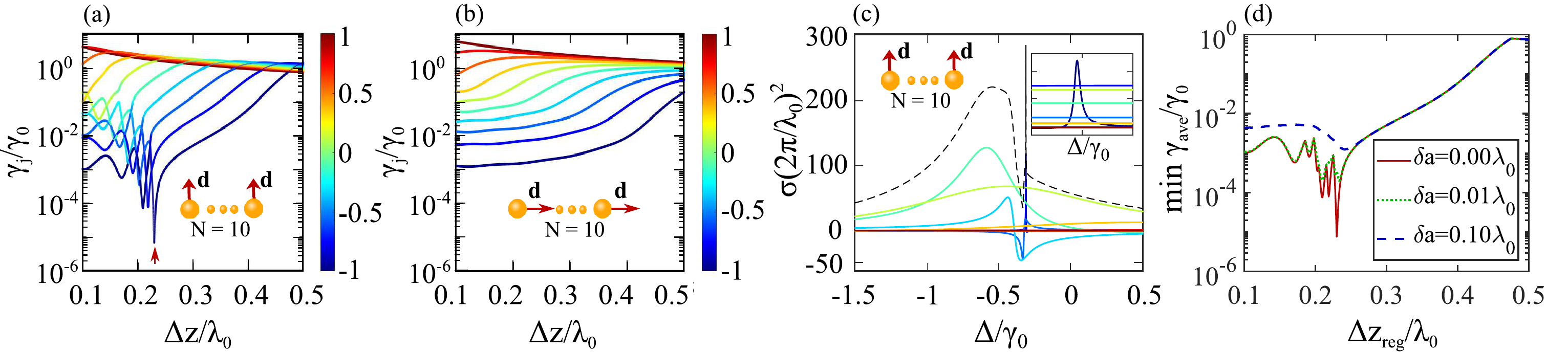}
	\caption{(Color online) Normalized spontaneous emission rates $\gamma_j$ of eigenstates as a function of the array period $\Delta z$ for $N = 10$ atoms. All atomic dipole moments are either purely transversal (a) or longitudinal (b). The color grade specifies the nearest-neighbor correlation function value, the bright red arrow points out the global minimum of the decay rate. (c) The total (black dashed) and the partial (color solid) cross sections ($\sigma_{tot}$ and $\sigma_j$) for the specific case of the transverse polarization and an array period $\Delta z \approx 0.23 \lambda_0$, shown by the red arrow in (a). The inset shows the region near the most subradiant state. (d) The effect of position disorder on a minimal collective decay rate $\gamma_j$ for an atomic array with a period $\Delta z_{\text{reg}}$. The  position of each atom is slightly fluctuating  according to a uniform distribution and it is plotted for different maximal deviations $\delta a$.}
	\label{10atGam}
\end{figure*}

In order to present the effect of subradiance in more details, we now analytically solve the problem of light scattering on an array of $N = 3$ two-level atoms, which has also been studied before in the context of superradiance \cite{FengPRA2013} and cooperative scatteting \cite{BettlesPRA2016}. 

The matrix of the level-shift operator Eq. \eqref{sigmavac} can be rewritten in terms of coupling constants, which are related to dipole-dipole interaction between atoms as follows:
\begin{eqnarray}
&\Sigma = \begin{pmatrix}
g_\text{self} & g_{1} & g_{2} \\
g_{1} & g_\text{self} & g_{1} \\
g_{2} & g_{1} & g_\text{self}
\end{pmatrix}, 
\end{eqnarray}
where $g_\text{self} = -i\gamma_{0}/2$, $g_{1}$ is the matrix element related to the interaction between the atoms being 1 period apart ($1-2, 2-3$), and $g_{2}$ for atoms 2 periods away from each other ($1-3$). In this context, the corresponding eigenvalues of this matrix can be easily found in the following forms:
\begin{eqnarray}
\lambda_1 &=& \dfrac{1}{2} \left(2 g_\text{self} + g_2 + \sqrt{8 g_{1}^{2} + g_2^{2}} \right), 
\nonumber\\
\lambda_2 &=& g_\text{self} - g_2,
\\
\lambda_3 &=& \frac{1}{2} \left( 2 g_\text{self} + g_2 - \sqrt{8 g_{1}^2 + g_2^ 2} \right). 
\nonumber
\label{eigenvalues3}
\end{eqnarray}
In Fig. \ref{3atGam} (a)-(b) we provide the spontaneous decay rates of these three states for the transversal and longitudinal photon polarizations respectively. In Fig. \ref{3atGam} (a) we can see that for a state corresponding to $\lambda_1$ (blue dashed line) it is possible to achieve a strong suppression of the emission rate for some array period $\Delta z$. Indeed, for the array of $N = 3$ atoms with $\Delta z \approx 0.14\lambda_0$ the imaginary part of $\lambda_1$ is more than an order of magnitude smaller than the linewidths of the two other states. 

Additionally, in order to characterize the collective effects of the system, we can introduce a \textit{nearest-neighbor correlation function} \cite{BettlesPRA2016}:
\begin{eqnarray}
\langle f_{i,i+1}^{j} \rangle = \dfrac{1}{N-1} \sum\limits_{i=1}^{N} \cos (\phi_{i+1}^{j} - \phi_{i}^{j}),
\label{nncf}
\end{eqnarray} 
as shown in Fig. \ref{3atGam} by color grading. Here a phase angle of the $i$-th component of the $j$-th eigenvector $\phi_{i}^{j} = \text{arg}[c_{i}^{j}]$ corresponds to a probability amplitude $c_i^j$ to have the excited atom $i$ for the eigenstate $j$. The function Eq. \eqref{nncf} gives the information about the phase correlation between neighbouring dipoles: it is equal to $+1$ for the neighboring dipoles with the same phases, and $-1$ for the neighboring dipoles with opposite phases \cite{BettlesPRA2016}. As one can see from Fig. \ref{3atGam}, this correlation function provides a useful information for a few atoms case, and allows distinguishing states with different symmetry. We also note that the state with the smallest value of $\langle f^{j}_{i,i+1} \rangle$ also possesses the smallest emission rate $\gamma_j$ due to the state symmetry; by further tuning $\Delta z$ it is possible to achieve a very small $\gamma_j$ as seen from \ref{3atGam} (a).

In Fig.\ref{3atGam} (c)-(d) we show the partial $\sigma_{j}(\Delta)$  and the total $\sigma_{\text{tot}}(\Delta)$ cross sections of the photon for two cases: when atoms have transverse (a, c) and longitudinal (b, d) dipole moment with respect to chain axis. The cross section profile of the subradiant state $\sigma_1(\Delta)$ is asymmetric due to the significant non-Lorenzian part $\sim f''_j$, as it appears in Eq. \eqref{ImTExp}.

In this simple and already studied example we have shown that there exists a specific interatomic spacing that allows to strongly suppress the emission rate of the system. In the next section we demonstrate what happens in an array with a larger number of atoms.

\subsection{Collective effects in an array of $N$ two-level atoms: highly subradiant states}

In this subsection we apply the developed formalism  to the case of $N$ two-level atoms in vacuum. Increasing the number of atoms leads to significant manifestation of collective effects. The difference between transverse and longitudinal cases becomes thereby more evident:
the transverse one shows a variety of highly subradiant states for different lattice periods $\Delta z$ as shown in Fig.~\ref{10atGam} (a). The difference in the behavior between  transverse and  longitudinal dipolar chains has been studied before in the context of optical properties of $1$D nanoparticle arrays \cite{WeberFord2004, BettlesPRA2016}.

The arrangement of atoms in $1$D chain with a sub-difractional period leads to a strong subradiance. Fig.~\ref{10atGam} (a)-(b) show the collective decay rates for an array of $N = 10$ atoms for various periods $\Delta z$. One can see that the strong subradiance appears only for transverse polarization. Moreover, this effect can be obtained for different atomic periods, as indicated by arrows in Fig.~\ref{10atGam} (a). Furthermore, from Fig.~\ref{10atGam} (b) it can be seen that interaction of an array of atoms with longitudinally polarized dipole moments leads to subradiance as well. But the dependence of the eigenvalues decay rate in this case is rather smooth and without any special features. To understand the difference of collective effects for different polarization in more details, let us compare the dipole-dipole coupling constants for these two cases:
\begin{eqnarray}
g^{\perp} &=& - \dfrac{3}{4} \hbar \gamma_0 e^{i k_0 \Delta z} \left[ \dfrac{1}{(k_0 \Delta z)} +  \dfrac{i}{(k_0 \Delta z)^2} - \dfrac{1}{(k_0 \Delta z)^3} \right], \nonumber\\
g^{||} &=& - \dfrac{3}{2} \hbar \gamma_0 e^{i k_0 \Delta z} \left[ - \dfrac{i}{(k_0 \Delta z)^2} + \dfrac{1}{(k_0 \Delta z)^3} \right]. 
\label{couplDD}
\end{eqnarray} 
 Now one can gain a physical intuition about the subradiance for different polarizations: there is no \textit{far-field} contribution in the dipole-dipole coupling constant in the case of the longitudinal polarization Eq. \eqref{couplDD}.
 Therefore, the strong subradiance results from an interplay between different types of fields: near-, intermediate-, and, importantly, far-fields. 
 
 \begin{figure}[b]
	\includegraphics[width=0.238\textwidth]{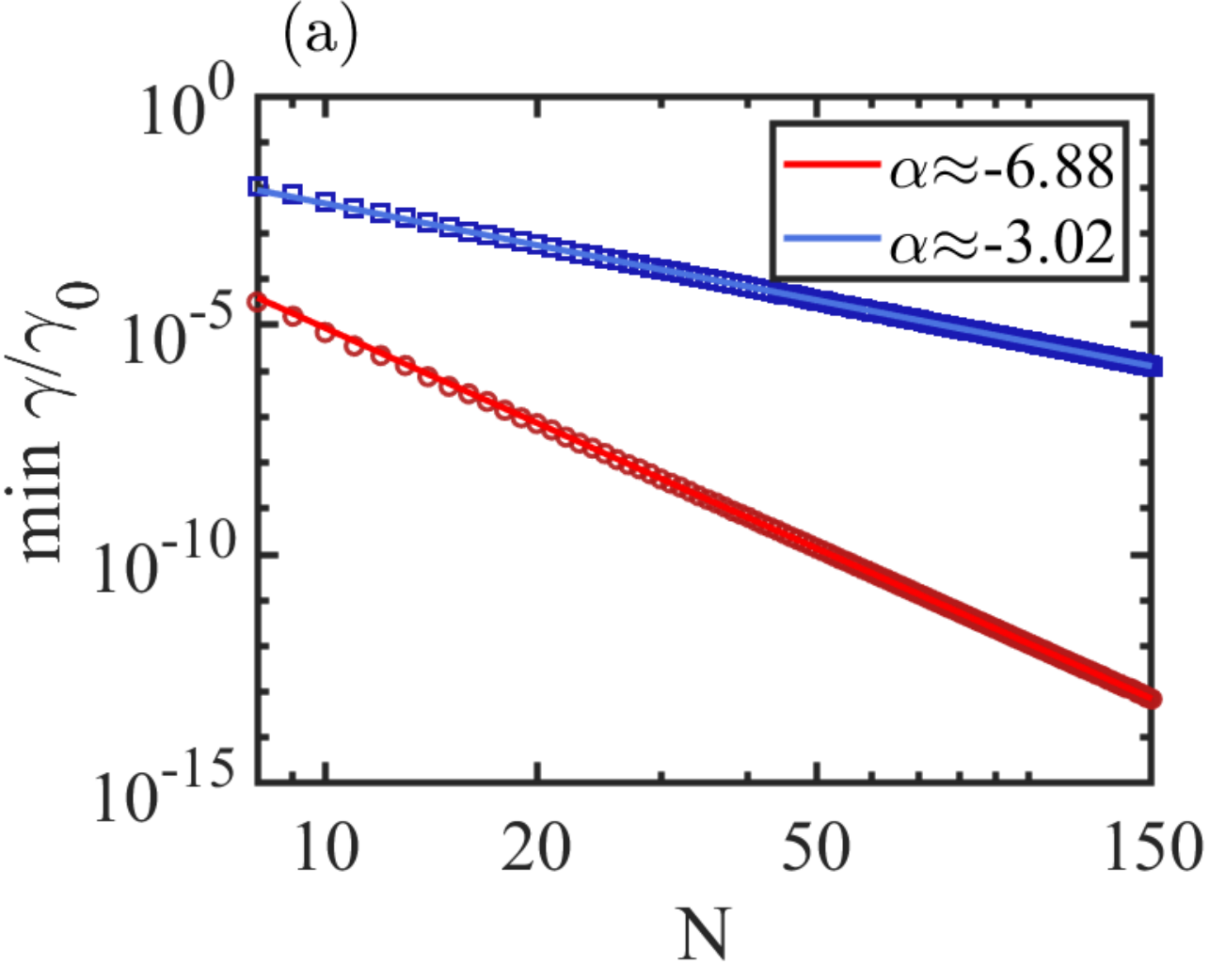} \includegraphics[width=0.238\textwidth]{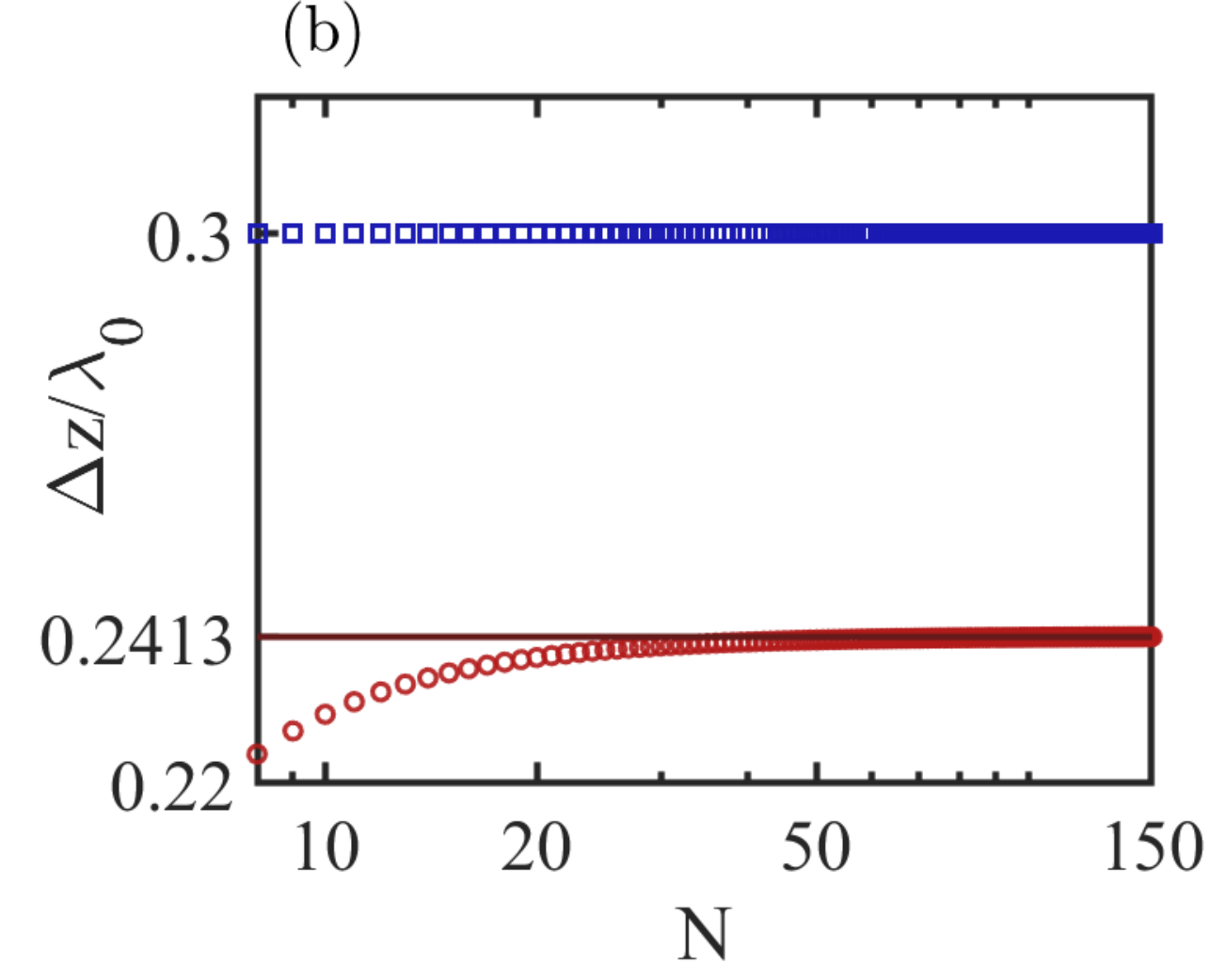}
	\caption{(Color online) (a) Collective emission rates for the most subradiant state as a function of the number of atoms $N$; blue open squares - for a regular period $\Delta z=0.3\lambda_0$, red open circles - for an optimal period $\Delta z_{\text{sub}}$, $\alpha$ specifies the characteristic scaling with the number of atoms: $\gamma_j \sim N^{\alpha}$, the corresponding fitting curves are specified by light blue and light red solid lines. In order to find $\alpha$ we used only data points for which $N \ge 20$. (b) The corresponding array periods $\Delta z$ versus number of atoms $N$ for which the subradiant state with the decay rate $\gamma_j$ can be achieved, note that the red open circles approach the value close to $\Delta z_{\text{sub}} \approx 0.24 \lambda_0$ for large $N$.}
	\label{minGvsN}
\end{figure}
 
 Furthermore, comparison of Fig.~\ref{3atGam} (a) and Fig.~\ref{10atGam} (a) reveals that the subradiant states appear at different atomic periods and with different spontaneous emission rates, which depend on the number of atoms $N$ in the chain. 

In Fig.~\ref{10atGam} (c) we also show the total cross section for the system period $\Delta z_{\text{sub}}$, which allows to achieve the minimal possible emission rate $\gamma_j$ (red arrow in Fig. \ref{10atGam} (a)). One can see that for $N=10$ the subradiant state appears in the spectrum as a sharp and asymmetric peak. It can be explained as a result of several overlapping resonances which contribute to the total cross section in this area (see inset of Fig. \ref{10atGam} (c)) leading to a large total cross section value in this spectral region.

Another feature of these dark states is their sensitivity to position disorder, when the atomic array is not perfectly periodic. In Fig. \ref{10atGam} (d) we show the dependence of the average minimal collective emission rate $\gamma_{ave}$ on the regular atomic array period introducing small fluctuation with uniform distribution and the maximal deviation $\delta a$. We see that even with small fluctuations $\delta a = 0.01\lambda_0$ in the atomic positions the resonances are almost smeared out, while a larger disorder induces a slight reduction of the emission rate in the range of regular system periods $0.20\lambda_0 < \Delta z_{\text{reg}} < 0.25 \lambda_0$.

\subsection{Emission rate scaling with atom number $N$}

It is also interesting to understand how does the emission rate of this highly subradiant state depend on the number of atoms $N$ in the chain. Previously in \cite{BlausteinOptExp2007}, it has been shown that in a sub-diffractional chain of dielectric particles the quality factor of the most bound modes scales as $\sim N^3$. Recently this question was also studied in the context of atomic chains, where the spontaneous emission rate for several most subradiant states decreases as $\sim N^{-3}$ \cite{TsoiPRA2008, ChangPRX2017} at least in some range of periods.

In our sub-diffractional atomic array with the lattice period $\Delta z_{\text{sub}}$ taken from Fig.~\ref{10atGam}, the value of the collective spontaneous emission rate scales as $\sim N^{-6.88}$, as shown in Fig. \ref{minGvsN} (a). A much faster decrease of the emission rate with number of atoms $N$ in comparison with the aforementioned studies \cite{BlausteinOptExp2007,TsoiPRA2008, ChangPRX2017} happens due to a proper choice of the system period $\Delta z_{\text{sub}}$, which allows to achieve a better destructive interference between scattering channels. We can see in Fig. \ref{minGvsN} (b), that this value is saturated to $\Delta z_{\text{sub}} \sim 0.24 \lambda_0$ for a large number of atoms $N$.

\begin{figure}[t]
	\includegraphics[width=0.485\textwidth]{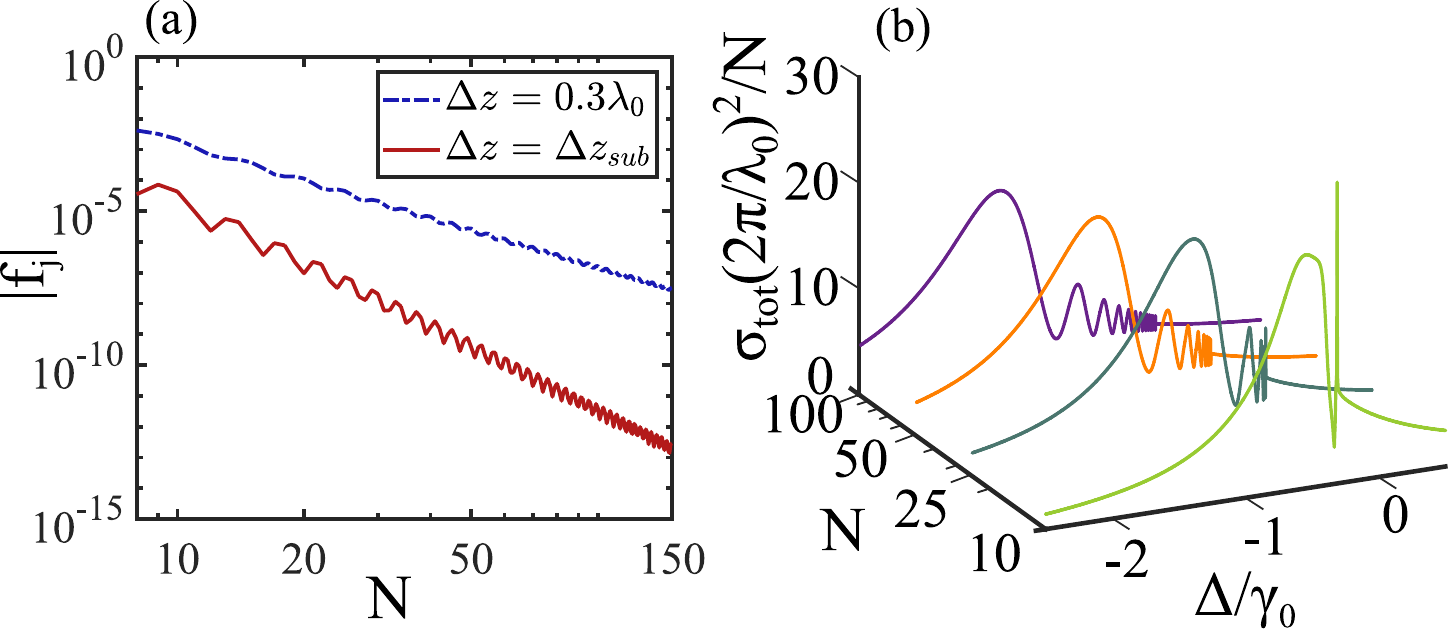}
	\caption{(Color online) (a) Dependence of the complex oscillator strength $|f_j|$ on the number of atoms $N$, blue dashed-dotted and red solid lines are for a fixed period ($\Delta z=0.3\lambda_0$) and $\Delta z_{\text{sub}}$ from Fig. 4(b), correspondingly. (b) The normalized total cross section $\sigma_\text{tot} (2\pi/\lambda_0)^2/N$ for different $N$. For each $N$ we choose the lattice period to be equal to $\Delta z_{\text{sub}}$.}
	\label{FPscalCS}
\end{figure}

Another physically important quantity is the oscillator strength amplitude of the corresponding collective eigenstate $f_j$. We can see from Fig. \ref{minGvsN} (a) and \ref{FPscalCS} (a) that $|f_j(N)|$ basically follows the same behavior as $\gamma_j$, but involves additional oscillations. This can be explained if one considers the overlap between the eigenstate $j$ with the $z$-propagating photon. In most cases, the "darker" the collective state is (and the smaller the corresponding decay rate $\gamma_j$ is), the smaller this overlap with the photon. Also we note that the distance between two neighboring local minima of $|f_j(N)|$ caused by the aforementioned oscillations is close to be $\Delta N \approx 4$. These oscillations induced by a bigger/smaller overlap between the atomic collective eigenstate and the $z$-propagating photon. In Fig. \ref{FPscalCS} (b) one can also see how does the total scattering cross section Eq. \eqref{ImTExp} for the system period $\Delta z_{\text{sub}}$ varies with atom number $N$: for sufficiently large $N$ subradiant states manifest themselves as a set of very sharp peaks, however, their relative contribution to spectrum becomes less pronounced.

\begin{figure}[b]
	\includegraphics[width=0.48\textwidth]{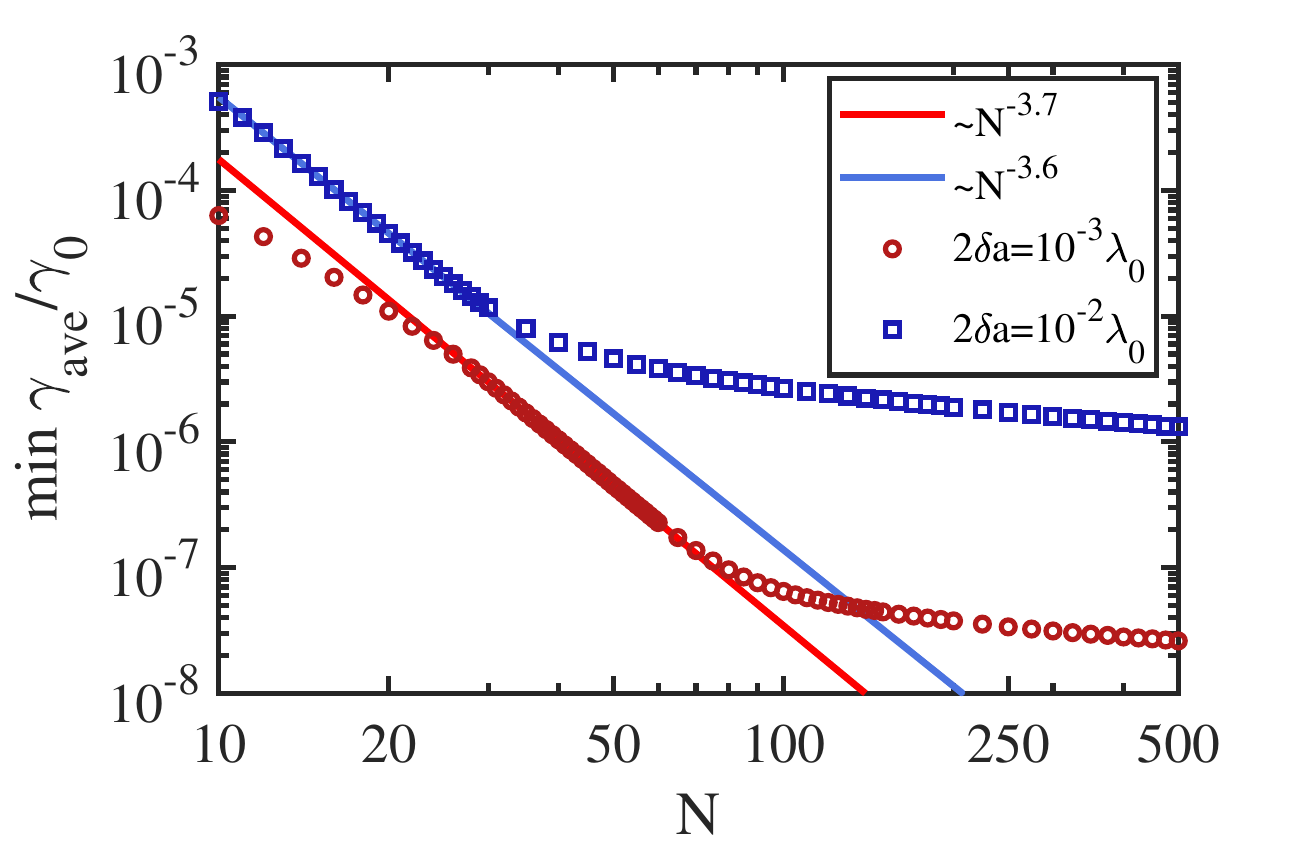}
	\caption{(Color online) Average minimal emission rate, corresponding to the most subradiant state, versus the atom number $N$ for the case of a uniform disorder of atomic $z$-positions. Two maximal position deviations are considered: $2 \delta a = 10^{-3}\lambda_0$ (open dark red circles), and $2 \delta a = 10^{-2}\lambda_0$ (open dark blue squares). Note that for both cases there are distinct regions of a $\sim N^{\alpha}$ behavior in the region of small $N$, where the linear fit in the double-log plot was performed, which is shown in solid bright red and solid bright blue lines, correspondingly.}
	\label{DisordScal}
\end{figure}

Finally, we address the scaling of the subradiant state emission rate in the presence of disorder in the atomic positions. In Fig.~\ref{DisordScal} we provide the emission rate scaling as a function of the number of atoms $N$ for two different disorders in the position of an atom $j$: $z_j = z_{\text{reg},j} + 2\delta a\cdot U(0,1)$. Here $z_{\text{reg},j} = (j-1)\Delta z_{\text{reg}}$ is the atomic position for a regular chain and $U(0,1)$  stands for a uniformly distributed pseudo random real number between $0$ and $1$.

From Fig.~\ref{DisordScal} one can see that disorder leads to a significantly slower decrease rate $\sim N^{-3.7}$ and this happens even for relatively small deviations from perfect periodicity $(2\delta a = 10^{-3}\lambda_0)$. However, increasing the number of atoms $N$ in the chain results in a transition to another regime, where a decrease of the emission rate is even slower. Numerical estimations show that in this region the scaling is on the order of $\sim N^{-0.3} - N^{-0.4}$. The main reason of this transition is related to disorder which induces localization of states. It can be estimated by calculating the Inverse Participation Ratio (IPR)  \cite{KramerRepProgPhys1993}: $IPR^{-1} = \sum_{j=1}^N \big|c_j^{(k)}\big|^4$, where $k$ is the label of a state of interest, $c_j^{(k)}$ is the probability amplitude that the atom $j$ is excited in a state $k$. For instance, for a positional disorder  $2\delta a = 10^{-2}\lambda_0$, and for a number of atoms $N = 20$ (the region of min $\gamma_{\text{ave}}/\gamma_0 \sim N^{-3.7}$ scaling), the average IPR of the corresponding subradiant state is equal to $IPR \approx 7.7$, while for $N = 200$ (where the scaling switches to min $\gamma_{\text{ave}}/\gamma_0 \sim N^{-0.36}$) $IPR \approx 9.7$. With this, one can see that for sufficiently large atom number $N \gg 1$, the subradiant states become strongly localized ($IPR \ll N$), and extension of the atomic chain does not modify the  $\text{min} \: \gamma_{\text{ave}}/\gamma_0$ significantly, contrary to the case of perfect periodicity. 

\section{Light scattering on an array of atoms trapped in the vicinity of an optical nanofiber}
So far we have studied the collective subradiance of atomic array in the vacuum. It is known that the light-atoms interaction can be significantly enhanced by placing the atomic system near a nanoscale object. Indeed, trapping atoms in the vicinity of an optical nanofiber dramatically changes the character of the atomic interaction and provides long-range dipole-dipole coupling between atoms not only via the vacuum, but also through the nanofiber guided mode. In this section we study modification of coupling effects coming from the scattering of the guided mode on an atomic chain trapped near the nanofiber surface (Fig. \ref{fiber}). 

\subsection{Theoretical framework of the light scattering process for an atomic array trapped near an optical nanofiber}

In this subsection we modify the developed formalism of light scattering in free space by introducing additional interaction via the nanofiber guided mode. Foremost, we need to modify the outer operators $\hat{V}$ in Eq. \eqref{Tmatgen}, which are responsible for absorption of the incoming guided photon and emission of the photon back into the same field mode. Furthermore, we are interested only in guided field modes of the outer operators $\hat{V}$.
However, the operator $\hat \Sigma(E_0)$ introduced in Eq. \eqref{sigmavac} for free space should include now all possible modes. 
The field subsystem in this configuration can be described using the quantization scheme proposed in \cite{Minogin2010}, where the quantized electric field of the nanofiber guided mode can be written as:
\begin{equation}
\mathbf{\hat E} (\mathbf{r}) = \sum_{\mu} \mathbf{E}_\mu(\mathbf{r}) \hat a_{\mu} + h.c.,
\end{equation}
where $\mathbf{E}_\mu$ is the electric field of the guided mode $\mu$ given by:
\begin{eqnarray}
\mathbf{E}_\mu (\mathbf{r}) = i \; \sqrt[]{\frac{2\pi \hbar \omega_\mu}{ \mathbb{L}}} \tilde{\mathbf E}_\mu(\rho, \phi) e^{if\beta_\mu z+im\phi}.
\end{eqnarray}
Here $\beta_\mu$ is the propagation constant, $\tilde{\mathbf E}_\mu(\rho, \phi)$ is the amplitude of the electric field, $\mathbb{L}$ is the quantization length, $f$ and $m$ define the direction of propagation and the mode angular momentum, respectively. The electric field is periodic in the $z$-direction with $\beta_l \mathbb{L}=2\pi l$, where $l$ is a positive integer.
And the electric field amplitude is normalized according to:
\begin{eqnarray}
\int\limits_{0}^{2\pi}\int\limits_{0}^{\infty}\ |\tilde{\mathbf E}_\mu(\rho, \phi)|^2 d\phi \rho d\rho = 1.
\end{eqnarray}
\begin{figure*}[t]
	\includegraphics[width=1\textwidth]{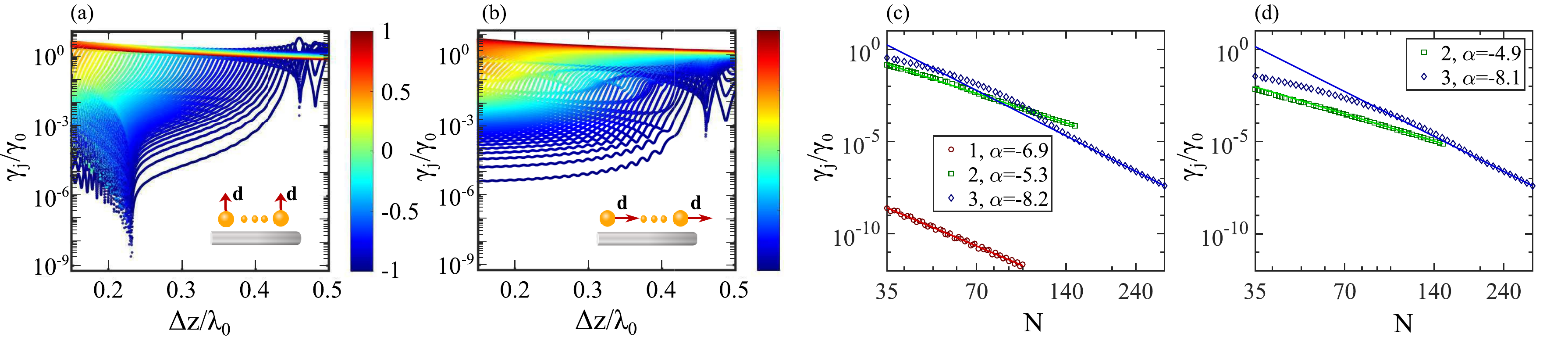}
	\caption{(Color online) (a) Spontaneous emission rates $\gamma_j/\gamma_0$ of the eigenstates $j$ for a periodic chain of $N = 75$ atoms placed near the nanofiber at a distance $\Delta\rho = \rho_c$ from the surface. The dipole moments of all atoms are aligned along the $\mathbf{e_\rho}$ direction. The nanofiber has  radius $\rho_c = 0.25\lambda_0$ and $\epsilon = 2.1$. (b) Same as in (a), but for dipole moments aligned along the axis $\mathbf{e_z}$. Color grade in (a)-(b) corresponds to values of the nearest-neighbor correlation function Eq. \eqref{nncf} for each collective eigenstate. (c)-(d): Local minima of the subradiant states in (a)-(b) with the number of atoms $N$. The solid lines correspond to linear approximations. Only the fundamental mode HE$_{11}$ was taken into account for these calculations.}
	\label{FiberCase}
\end{figure*}

For simplicity, we assume that all atoms are located at the same distance $\Delta\rho$ from the nanofiber surface. All matrix elements of the outer operator $\hat V$ have, therefore, the same absolute values and differ only by phases. The outer operators $\hat V$ presented in the matrix $T$ in Eq. \eqref{Tmatgen} can finally be written as:
\begin{eqnarray}
\langle e_a,\{0\}| && \hat V |g_a, 1_{\mu^\prime}\rangle = 
\nonumber\\
&& - i \left( \mathbf{d}_{a,eg}\cdot \mathbf{\tilde {E}_{\mu^{\prime}}} (\rho_a, \phi_a)\right) \sqrt{\dfrac{2\pi \hbar \omega_{\mu^{\prime}}}{\mathbb{L}}} e^{i\beta_{\mu^{\prime}}z_a + i \phi_a} 
\nonumber\\
\langle g_b, 1_{\mu^{\prime\prime}}| && \hat V | e_b, \{0\}\rangle = 
\\
&& i \left( \mathbf{\tilde{E}^{*}_{\mu^{\prime\prime}}}(\rho_b, \phi_b)\cdot \mathbf{d}_{b, ge} \right)
\sqrt{\dfrac{2\pi \hbar \omega_{\mu^{\prime\prime}}}{\mathbb{L}}}
e^{-i\beta_{\mu^{\prime\prime}}z_b - i \phi_b}
\nonumber
\label{OuterVGuided}
\end{eqnarray}
where $\beta_{\mu}$ is the propagation constant of the guided mode $\mu$.

At the next step, we calculate the matrix elements of the operator $\hat{\Sigma}$ in the presence of the nanofiber. The theoretical description of excitation transfer between atoms through the radiation into vacuum and nanofiber guided modes was done in \cite{Gruner1996}. Using this formalism, the Hamiltonian of our system can be written as:
\begin{eqnarray}
&&\hat{H}_0 = \sum\limits_n \hbar \omega_0 \hat{\sigma}^{+}_n \hat{\sigma}^{-}_n + \int d\mathbf{r^{\prime}} \int\limits_0^{\infty} d\omega^{\prime} \hbar \omega^{\prime} \hat{\mathbf{f}}^\dagger (\mathbf{r^{\prime}},\omega^{\prime}) \hat{\mathbf{f}} (\mathbf{r^{\prime}},\omega^{\prime}),
\nonumber \\ 
&&\hat{V} = -\sum\limits_n \hat {\mathbf d}_{n} \hat {\mathbf E}(\mathbf{r}_n),
\end{eqnarray}
where $\omega_0$ is the atomic transition frequency. $ \hat{\mathbf E}(\mathbf{r}_n)$ is the total electric field and $\hat{\mathbf{f}} (\mathbf{r^{\prime}}, \omega^{\prime}), \hat{\mathbf{f}}^{\dagger} (\mathbf{r^{\prime}}, \omega^{\prime})$ are the bosonic vector local-field operators, which obey the following commutation relations:
\begin{eqnarray}
\left[\hat{f}_i(\mathbf{r^{\prime}}, \omega^{\prime}), \hat{f}^{\dagger}_k(\mathbf{r}, \omega)\right] &=& \delta_{ik}\cdot\delta(\mathbf{r^{\prime}} - \mathbf{r})\cdot \delta(\omega^{\prime} - \omega),
\nonumber\\
\left[\hat {f}_i(\mathbf{r^{\prime}}, \omega^{\prime}), \hat{f}_k(\mathbf{r}, \omega) \right] &=& 0
\end{eqnarray}
The positive-frequency part of the total electric field has the following form:
\begin{eqnarray}
&&\hat{\mathbf{E}}^{+}(\mathbf{r}) =
\\
&& i \; \sqrt[]{4 \hbar} \int d\mathbf{r^{\prime}} \int\limits_{0}^{\infty} d\omega^{\prime}\frac{{\omega^{\prime}}^2}{c^2} \sqrt{\epsilon_I (\mathbf{r^{\prime}}, \omega^{\prime})} \mathbf{G}(\mathbf{r}, \mathbf{r^{\prime}}, \omega^{\prime})\cdot \hat{\mathbf{f}}(\mathbf{r^{\prime}}, \omega^{\prime}),
\nonumber
\end{eqnarray}
where $\epsilon_I(\mathbf{r}^{\prime}, \omega^{\prime})$ is the imaginary part of the dielectric permittivity of the media, $\mathbf{G}(\mathbf{r}, \mathbf{r}^{\prime}, \omega^{\prime})$ is the classical Green's tensor of the electric field. In the presence of  {the} optical nanofiber the Green's tensor can be expanded into:
\begin{eqnarray}
\label{Greens}
\mathbf{G}(\mathbf{r}, \mathbf{r}^{\prime}, \omega)  = \mathbf{G}_0(\mathbf{r}, \mathbf{r}^{\prime}, \omega) + \mathbf{G}_s(\mathbf{r}, \mathbf{r}^{\prime}, \omega),
\end{eqnarray}
where $\mathbf{G}_0$ is the vacuum Green's tensor, and $\mathbf{G}_s$ is the Green's tensor corresponding  to light scattering from the nanofiber. The scattering term of the Green's tensor can be expanded into the Vector Wave Functions and the details of these calculations are given in Appendix A.
In the lowest non-vanishing order, the matrix elements of the level-shift operator can be written as:
\begin{eqnarray}
\langle f| \hat \Sigma (E) |i\rangle =\sum_{|\alpha \rangle , |\beta \rangle} \langle f|\hat V|\alpha \rangle \langle \alpha | \frac{1}{E - \hat H_0 + i\eta}|\beta \rangle \langle \beta |\hat V|i\rangle,  \nonumber \\
\label{LevelShiftFiber}
\end{eqnarray}
where $|i\rangle$ and $|f\rangle$ are the initial and final states of the system, respectively; $|\alpha \rangle, |\beta \rangle$ are the two possible intermediate states with a single elementary excitation for the field subsystem. Both atoms are either in the excited or the ground state:
\begin{eqnarray}
|e_{n},e_{m}\rangle \times \hat {\mathbf{f}}^\dagger (\mathbf{r^\prime},\omega^\prime)|\{0\}\rangle, \nonumber 
\\
|g_{n},g_{m}\rangle \times \hat {\mathbf{f}}^\dagger (\mathbf{r^\prime},\omega^\prime)|\{0\}\rangle.
\end{eqnarray}

Further details on the derivation of the matrix elements of the level-shift operator Eq. \eqref{LevelShiftFiber} can be found in \cite{Dung2002} and here we provide only the final expression:
\begin{equation}
\langle f| \hat \Sigma(E) |i\rangle = - 4\pi \frac{\omega_0^2}{c^2} \mathbf{d}_{m, eg} \cdot \mathbf{ G}(\mathbf{r}_m,\mathbf{r}_n,\omega_0)\mathbf{d}_{n, ge}.
\label{SigmaOp}
\end{equation}

The matrix $\hat{\Sigma}(E)$ can be found using Eq. \eqref{SigmaOp}. Note, that the scattering matrix Eq. \eqref{Smatgen} is also valid in the presence of the nanofiber. In the field quantization scheme that we use here one should include summation over final states into Eq. \eqref{Smatgen}, go into the limit $\mathbb{L} \to \infty$, which means that the propagation constant $\beta$ can now be continuous. This limit is equivalent to $\sum\limits_{n_{\beta}} \to \frac{\mathbb{L}}{2 \pi} \int\limits_0^\infty \dfrac{d\beta}{d\omega} d\omega$.
In the end we seek the expression for the scattering matrix: 
\begin{equation}
S_{f',p';fp} = \delta_{f',p';fp} - i\frac{\mathbb{L}}{c \hbar \cdot dk/d\beta}T_{f',p';fp}(E),
\end{equation} 
which can be also found in \cite{Pivovarov2018}.
\subsection{$S$ matrix in the eigenstate picture}
Let us now analyze the scattering process considering the eigenstates of the system. For this, we need to diagonalize the matrix $\Sigma(\omega_0)$, which is responsible for the coupling of different atomic states through the field modes. We should mention that in the case of a reciprocal problem $\Sigma$ is a symmetric matrix, while in a non-reciprocal one (when, for example, a magnetic field is applied to separate the $\sigma^{+}$ and $\sigma^{-}$ transitions making only one of them active at a given frequency) $\Sigma$ is not symmetric anymore, and it has different right and left eigenvectors. However, the diagonalization of the problem can be simply done by choosing the right eigenvectors, for instance:
\begin{eqnarray}
&&\Sigma(E_0) v^{(r)}_j = \lambda_j v^{(r)}_j,
\nonumber\\
&&(S^{(r)})^{-1} \dfrac{1}{I\hbar \Delta - \Sigma(E_0)} S^{(r)} = \dfrac{1}{I\hbar \Delta - \Lambda},
\end{eqnarray}
where the matrix $S^{(r)}$ is the transformation matrix to the eigenspace whose columns are the right eigenvectors $v^{(r)}$ of $\Sigma(E_0)$, $\Delta$ is the photon detuning from the atomic resonance, and $\Lambda$ is a diagonal matrix having the corresponding eigenvalues $\lambda_j$ as its entries. 

In order to simplify the final expression for the scattering matrix, we can express the product of two outer $\hat V$ matrix elements Eq. \eqref{OuterVGuided} using the known relations for the spontaneous emission rate into the forward-propagating guided modes:
\begin{multline}
\gamma^{(f)}_{wg} = \sum\limits_{m} \dfrac{3 \pi |\mathbf{n}_{eg} \mathbf{\tilde{E}}_{f=+1,m}|^2  d\beta/dk}{2 k_0^2} \gamma_0 = \\
\sum\limits_{m} \dfrac{2 \pi |\mathbf{d}_{eg} \mathbf{\tilde{E}}_{f=+1,m}|^2 k_0 \cdot d\beta/dk }{\hbar}
\label{gamguided}
\end{multline}
In this special symmetry we can push the coupling constant to the forward-propagating guided mode $\gamma_{wg}^{(f)}$ outside of the sum over the eigenstates and finally rewrite the $S$ matrix element corresponding to forward scattering as:
\begin{equation}
S_{ii} = 1 - i \hbar \gamma_{wg}^{(f)} \sum\limits_j \dfrac{f_j^{(t)}}{\hbar \Delta - \lambda_j}, 
\label{ST}
\end{equation}
which has a form similar to Eq. \eqref{sigmatotexp} with $f_j^{(t)}$ being complex-valued constants. We observe that indeed considering equal coupling strengths for all of the atoms is clear: in this case coefficients $f_{j}^{(t)}$ are dimensionless and carry information only about the phase. However, it is possible to rewrite it for a general situation, but the meaning of $f_{j}^{(t)}$ will be slightly different and it will take into account the couplings of individual atoms to the guided mode.

The light scattering in an one-dimensional configuration can be characterized by a transmission coefficient $t = |S_{ii}|^2$, which can be written in the following form, as shown in Appendix B:
\begin{eqnarray}
|S_{ii}|^2 &=& 1 + 2 \hbar \gamma_{wg}^{(f)} \sum\limits_{j=1}^{N} \left[   \dfrac{\eta_j^{(t)} \lambda_j'' + \xi_j^{(t)}\left(\hbar \Delta - \lambda_j'\right) }{(\hbar \Delta - \lambda_j')^2 + \lambda_j''^2} \right], 
\nonumber\\
\eta_j^{(t)} &=& f_j^{(t)'} - \sum\limits_{i=1}^{N} \hbar \gamma_{wg}^{(f)} \text{Im} \left[ \dfrac{f_j^{(t)} (f_i^{(t)})^*}{\lambda_j - \lambda_i^*}\right],
\nonumber\\
\xi_{j}^{(t)} &=&  f_j^{(t)''} + \sum\limits_{i=1}^{N} \hbar \gamma_{wg}^{(f)} \text{Re} \left[ \dfrac{f_j^{(t)} (f_i^{(t)})^*}{\lambda_j - \lambda_i^*} \right]. 
\label{TExp}
\end{eqnarray}
One can see that the transmission, as the cross-section for the vacuum case Eq. \eqref{ImTExp}, includes Lorenzian and non-Lorenzian terms. However, the respective dimensionless coefficients $\xi_j^{(t)}, \eta_j^{(t)}$ differ: apart from $f_j^{(t)'}$, $f_j^{(t)''}$ there are also terms expressed through $\frac{f_j^{(t)} (f_i^{(t)})^*}{\lambda_j - \lambda_i}$, which can be associated with the interference of $i$ and $j$ resonances.

Similarly, we can expand the reflection coefficient and the corresponding reflectance as:
\begin{eqnarray}
S_{bf} &=& - i \hbar \sqrt{\gamma_{wg}^{(f)} \gamma_{wg}^{(b)}} \sum\limits_{j} \dfrac{f_{j}^{(r)}}{\hbar \Delta - \lambda_j} \nonumber\\
| S_{bf} |^2 &=& \sum\limits_{i, j} \hbar^2 \gamma_{wg}^{(f)} \gamma_{wg}^{(b)} \dfrac{f_j^{(r)}(f_i^{(r)})^*}{(\hbar \Delta - \lambda_j) (\hbar \Delta - \lambda_i^*)} = 
\nonumber\\
&& 2 \hbar \gamma_{wg}^{(f)} \sum\limits_{j=1}^{N} \dfrac{ \left( \eta_{j}^{(r)} \lambda_j'' + \xi_{j}^{(r)} (\hbar \Delta - \lambda_j') \right)}{(\hbar \Delta - \lambda_j')^2 + \lambda_j''^2},
\nonumber\\
\eta_j^{(r)} &=& - \hbar \gamma_{wg}^{(b)} \: \text{Im} \: \sum\limits_{i=1}^{N} \left[ \dfrac{f_j^{(r)} (f_i^{(r)})^*}{\lambda_j - \lambda_i^*} \right], \nonumber\\
\xi_{j}^{(r)} &=& \hbar \gamma_{wg}^{(b)} \: \text{Re} \: \sum\limits_{i=1}^{N} \left[ \dfrac{f_j^{(r)}(f_i^{(r)})^*}{\lambda_j - \lambda_i^*}\right].
\label{RExp}
\end{eqnarray} 
We analyze the subradiance in a presence of a nanoscale waveguide in next subsection.

\subsection{Highly subradiant states and atom number scaling in the presence of a nanofiber}

We now apply the developed formalism to the perfectly periodic $1$D array of atoms trapped near a nanofiber. Similarly to the free-space configuration,
in Fig. \ref{FiberCase} we show the spontaneous emission rate of each eigenstate of the periodic chain of $N = 75$ atoms into the fundamental nanofiber mode HE$_{11}$, as well as the dependence of the spontaneous emission minima on the number of atoms $N$. 

\begin{figure}[t]
	\includegraphics[width=0.48\textwidth]{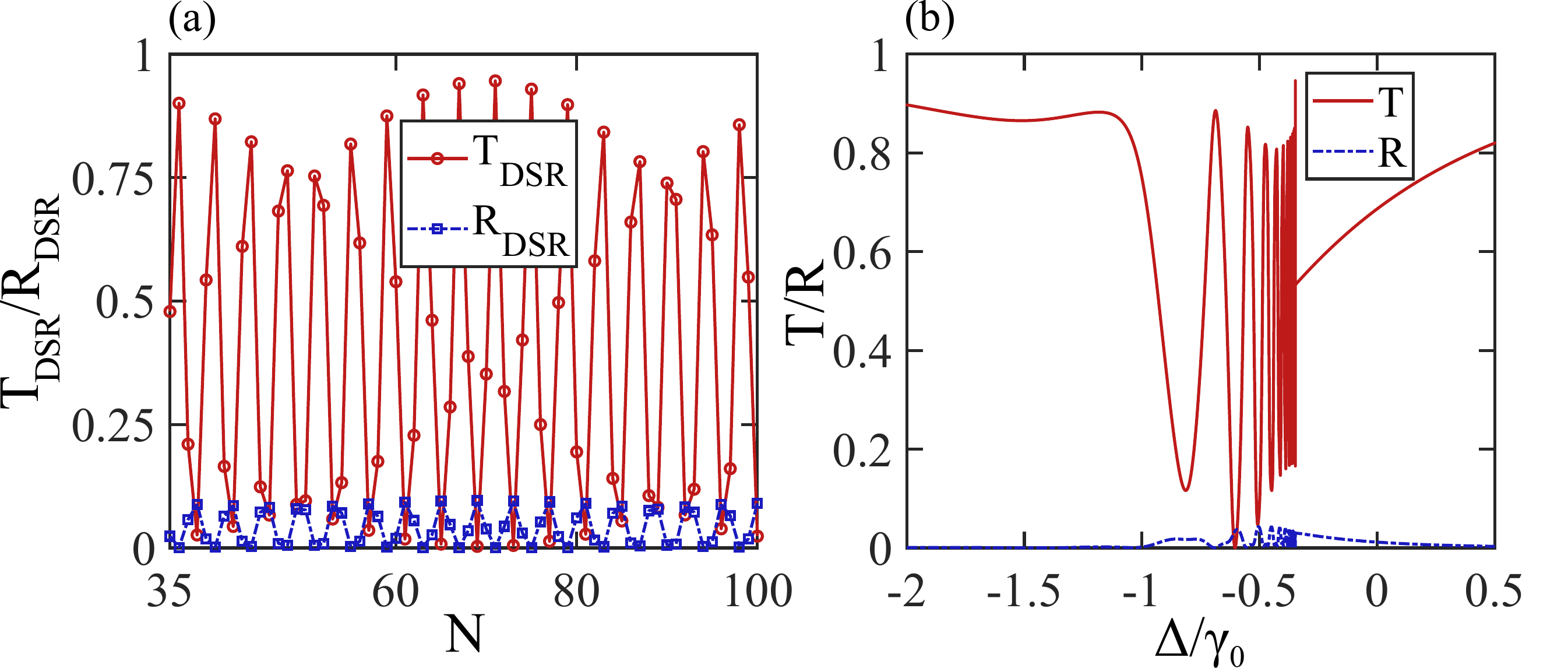}
	\caption{(Color online) (a) Transmission (red circles, solid line) and reflection (blue squares, dashed-dotted line) of guided light in a $1$D array of $N$ trapped atoms with a lattice period $\Delta z = \Delta z_{\text{sub}}$ corresponding to the dark state. (b) Transmission (red solid line) and reflection (blue dashed-dotted line) spectra for $N = 71$, which corresponds to the highest $T_\text{DSR}$ in (a). All fiber parameters are the same as for Fig. \ref{FiberCase}.}
	\label{1stResSpec}
\end{figure}

The third subradiant resonance at $\Delta z \approx 0.48 \lambda_0$ can be explained as a result of interference between the two interaction channels: the vacuum modes and the guided mode. In Fig. \ref{FiberCase} (c)-(d) we show the scaling of these minima with the number of atoms $N$. One can see, that for sufficiently large $N$ all three curves follow to $\sim N^{\alpha}$ dependency. 

\subsection{Subradiant states in the transmission and reflection spectra.}

In the context of the specific $1$D configuration of the system, it is interesting to study the transmission and reflection coefficients at the subradiant resonance condition. 

In Fig.~\ref{1stResSpec} (a) we show the dependence of the transmission and the reflection coefficients on the number of trapped atoms $N$ for the first subradiant state. One can see that these coefficients have oscillating behavior making the system either transparent with $T \ge 0.75$ and $R \sim 0$ or opaque with $T \le 0.10$ and $R \approx 0.10$. A corresponding spectrum in the first subradiant state range of $\Delta z$ is shown in Fig.~\ref{1stResSpec} (b) for $N = 71$ atoms, where one can see many distinct subradiant states and a very sharp resonance with $T \approx 0.90$. The nature of such oscillations in $T_\text{DSR}/R_\text{DSR}$ is of the same nature as was discussed for the scattering of a photon on a transverse chain in the vacuum: it appears due to oscillating value of the overlap between the atomic eigenstate and the photon, and these oscillations have the same distance between the local minima of $\Delta N = 4$.

\begin{figure}[!t]
	\includegraphics[width=0.48\textwidth]{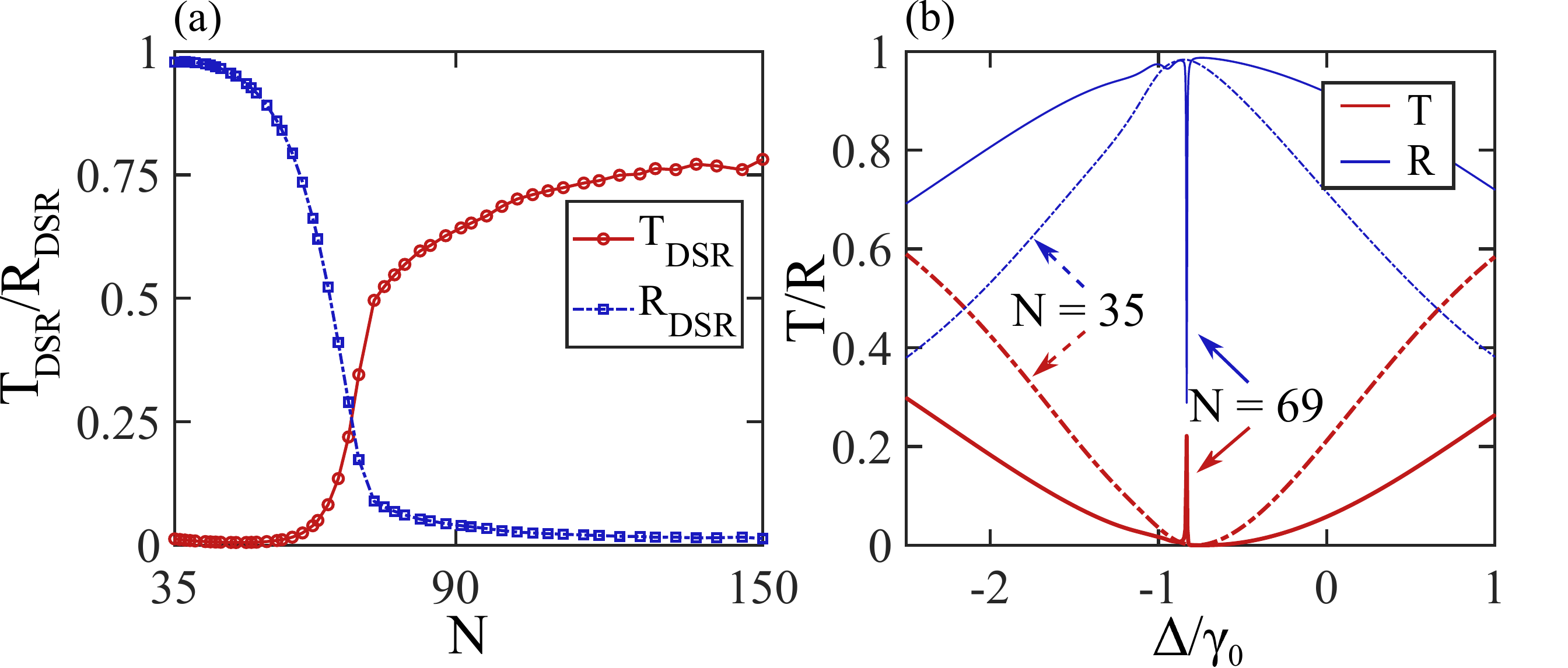}
	\caption{(Color online) (a) Same as in Fig. \ref{1stResSpec} (a), but at the second subradiant resonance condition. (b) Transmission (red) and reflection (blue) spectra for $N = 35$ (dashed-dotted) and $N = 69$ (solid) atoms.  All other relevant parameters are the same as for Fig. \ref{FiberCase}.}
	\label{2ndResSpec}
\end{figure}

The transmission and reflection for the second subradiant state that appears at the Bragg resonance condition for the fundamental guided mode are shown in Fig.~\ref{2ndResSpec}. One can see that at the second subradiant state the atom-atom interaction has different behavior. Effectively, one can say that the total spectrum mainly consists of the superradiant and a very long-lived subradiant state $T = |S_{ii}|^2 = 1 + T^\text{BS} + T^\text{DS}$, where BS/DS correspond to Bright/Dark State \cite{PlankensteinerPRL2017}. Foremore, increasing the number of atoms $N$ makes this dark state more distinguishable and leads to an increase of the transmission and reduction of the reflection, see Fig.~\ref{2ndResSpec} (a). At the same time, the system can be purely transparent or purely reflective for a small number of atoms at this resonance condition and can have a very narrow transparency window near $\Delta/\gamma_0 \approx -1$ for the large number of atoms, as it is shown in Fig.~\ref{2ndResSpec} (b).

\begin{figure}[t]
	\includegraphics[width=0.48\textwidth]{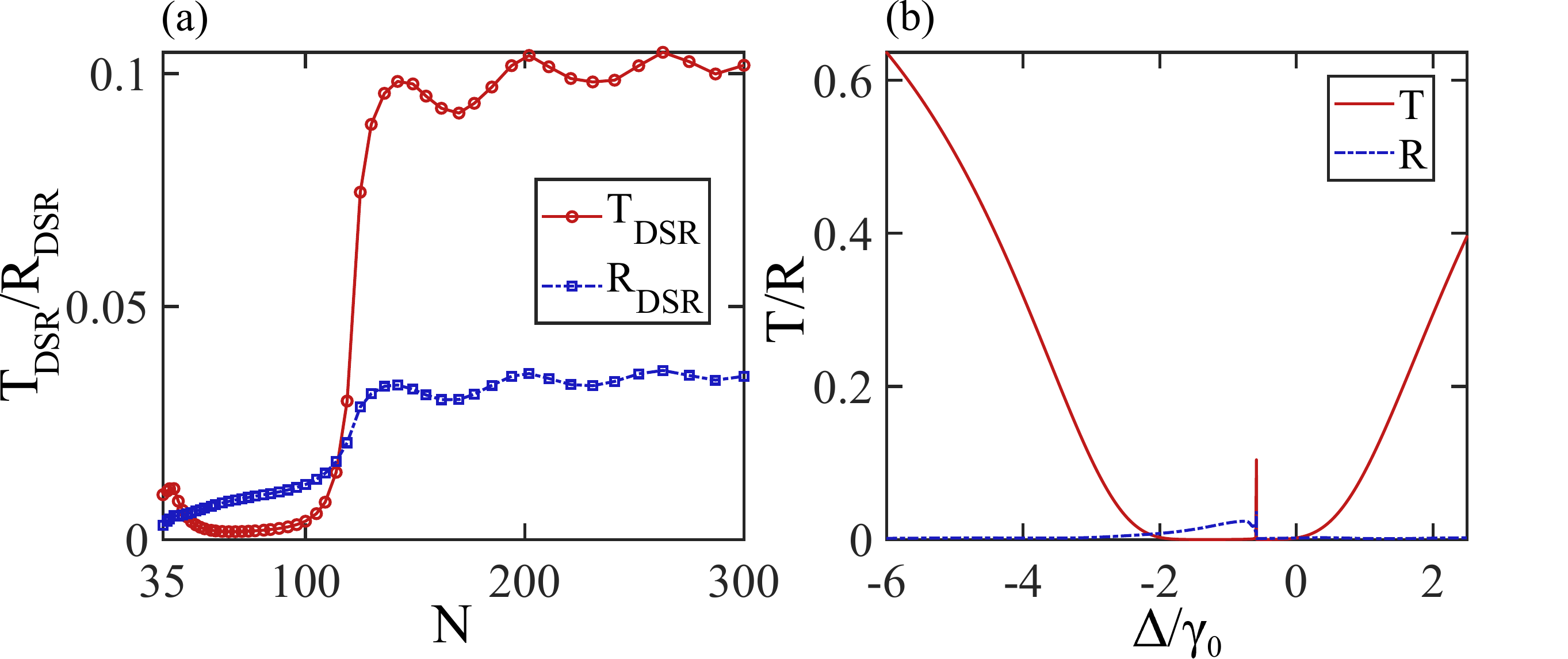}
	\caption{(Color online) (a) Same as in Fig.~\ref{1stResSpec} (a), but for the third subradiant resonance condition. (b) Transmission (red solid line) and reflection (blue dashed-dotted line) spectra for $N=202$. All other relevant parameters are the same as for Fig.~\ref{FiberCase}.}
	\label{3rdResSpec}
\end{figure}

The third subradiant state, which is the result of interaction between the vacuum and the nanofiber guided modes does not show any particular interesting behavior: both the transmission and the reflection are small (less than $0.1$) for the considered number of atoms $N$, as shown in Fig.~\ref{3rdResSpec} (a). At the same time, the subradiant state manifests itself as a sharp resonance with a small amplitude of both the transmission $T$ and the reflection $R$, see Fig. \ref{3rdResSpec} (b).


\section{Conclusion}

In conclusion, we have studied the subradiant collective states for a periodic array of two-level atoms with a given dipole moment transition in the sub-diffractional regime. We considered the atomic array both in free-space and trapped in the vicinity of an optical nanofiber. Trapping atoms with transversal dipole moments in an one-dimensional array with specific lattice periods $\Delta z$ provides a significant reduction of the collective emission rate, the emission rate can be decreased further by taking a bigger number of atoms $N$. Importantly, we have shown that this dependency on the number of atoms is $\sim N^{-6}-N^{-7}$, unlike the known so far $\sim N^{-3}$ scaling. We have found that the corresponding period has an asymptotic value  $\Delta z \approx 0.24\lambda_0$ for a  large number of atoms $N$ in the vacuum. In addition, we studied the scaling of the collective emission rate in the presence of positional disorder along the chain. We have shown that the introduction of disorder in the system leads to a slower decrease of the emission rate up to $\sim N^{-3.6}$. There is also a transition to a significantly slower decrease rate after a certain number of atoms $N$, which happens due to disorder-induced localization of states. 

Moreover, we have studied the subradiant states when the interaction between the atoms is also provided by the fundamental guided mode of an optical nanofiber. We showed that in this scenario there are extremely subradiant states similar to those mentioned before for the vacuum case and they affect the optical properties of the system like transmittance and reflectance leading to the presence of very sharp peaks in the spectra. We found that on the corresponding resonance frequency the system becomes either highly transparent or opaque depending upon the number of atoms $N$.

There are also two other types of subradiant states appear in a presence of a nanofiber. These states are present for both transverse and longitudinal case: one can be observed on the first Bragg resonance for the nanofiber guided mode and the other one is the result of an interplay between the vacuum and guided mode interaction channels. The former one allows the subradiant state on the first Bragg resonance to be visible in the transmission/reflection spectra opening a very narrow window of a partial transparency for sufficiently large $N$. The latter is weakly pronounced in the spectra at least for the considered set of parameter.

Extremely subradiant states studied in this work may find applications in both atomic optics and quantum  information science in relation to the problem of quantum memory for instance \cite{ChangPRX2017}. We also want to note that such states might be studied not only in the context of cold atoms physics, but also nanophotonics: e.g. $1$D arrays of dielectric or plasmonic nanoparticles. In this field of research long-lived states might be exploited in order to create tunable discrete waveguides, where the optical properties of an overall system are defined by characteristics of individual elements and their arrangement.

\section*{Acknowledgement}

D.F.K. acknowledges the support from the Embassy of France in Russia via the Ostrogradski scholarship, and the support from Basis Foundation. This work was supported by the European Research Council (Starting Grant HybridNet), Sorbonne Université (PERSU program), the French National Research Agency (NanoStrong project), the Ile-de-France DIM Sirteq, and by Russian Science Foundation (project \# 19-72-10129). A.S.S. is supported by the European Union (EU) (Marie Curie fellowship  NanoArray).

\appendix

\section{}
\label{App A}
The classical electromagnetic Green's tensor of our system can be found from the vector Helmholtz equation:  
\begin{eqnarray}
\left[- \frac{\omega^2}{c^2}\veps(\mathbf{r},\omega) + \mathbf{\nabla} \times \mathbf{\nabla} \times  \right] \mathbf{G}(\mathbf{r},\mathbf{r}^\prime,\omega) = \mathbf{I}\delta(\mathbf{r}-\mathbf{r}^\prime),
\nonumber\\
\label{A1}
\end{eqnarray}
where $\veps(\mathbf{r},\omega)$ is the complex dielectric function and $\mathbf{I}$ is the unit dyad. In our case we consider a dielectic cylindrical waveguide of radius $\rho_c$ and dielectric permittivity $\veps$ being constant inside the cylinder. To find the solution we apply the scattering superposition method \cite{Chew1999, Tai1994}, which allows to expand the Green's tensor into the homogeneous and inhomogeneous terms:
\begin{equation}
\mathbf{G}(\mathbf{r},\mathbf{r}^\prime,\omega) = \mathbf{G}_0(\mathbf{r},\mathbf{r}^\prime,\omega) + \mathbf{G}_s(\mathbf{r},\mathbf{r}^\prime,\omega).
\end{equation}
As soon as we consider dielectric particles in the vicinity of the waveguide, so that $\mathbf{r}, \mathbf{r}^\prime$ are outside the cylinder, the homogeneous term is always present and describes the field directly generated at the field point $\mathbf{r}$ by the source placed at the point $\mathbf{r}^\prime$. This term can be obtained analytically from the Green tensor written in cartesian coordinates using the transformation from cartesian to cylindrical coordinates $\mathbf{S}(\phi)\mathbf{G}_{0}^{\text{Cart}}(\mathbf{r},\mathbf{r}^\prime,\omega)\mathbf{S}^T(\phi')$, where $\mathbf{G}_{0}^{\text{Cart}}$ has an analytic expression \cite{Novotny2012} and is given by 
\begin{eqnarray}
\mathbf{G}_{0}^{\text{Cart}}(\mathbf{r}, \mathbf{r^\prime}, \omega) = \left( \mathbf{I} + \frac{1}{k^2}\mathbf{\nabla} \otimes \mathbf{\nabla}\right)G_0(\mathbf{r},\mathbf{r}^\prime,\omega),
\nonumber\\
\end{eqnarray}
here $G_0(\mathbf{r},\mathbf{r}^\prime,\omega)$ is the  Green's function of the scalar Helmholtz equation.

The scattering term can be calculated via the integral representation of the homogeneous part. To obtain this representation we apply the method of Vector Wave Functions (VWF) explained in details in Ref.~\cite{Chew1999, Tai1994}, here we cover only the basic ideas and provide the final expressions.
To find the solution of the vector Helmholtz equation \eqref{A1}, we introduce the scalar Helmholtz equation and  the solution of this equation in the cylindrical coordinates:
\begin{eqnarray}
&& \nabla^2\phi(\mathbf{k},\mathbf{r}) + k^2\phi(\mathbf{k},\mathbf{r}) = 0, 
\nonumber\\
&& \phi_n(k_{z},\mathbf{r})=J_n(k_\rho \rho)e^{in\theta+ik_zz},
\end{eqnarray}
here $J_n(x)$ is the Bessel function of the first kind, ${\mathbf{r} = (\rho,\theta,z)}$ are the cylindrical coordinates and $k_\rho$, $k_z$ are the projections of the wavevector $\mathbf{k}$.
The solution of the vector Helmholtz equation may be written in terms of the following vector wavefunctions:
\begin{eqnarray}
\mathbf{M}_n(k_z,\mathbf{r}) &=& \mathbf{\nabla} \times [\phi_n(k_z,\mathbf{r})\mathbf{e_z}] 
\nonumber\\
\mathbf{N}_n(k_z,\mathbf{r}) &=& 
\frac{1}{k}\mathbf{\nabla } \times \mathbf{M}_n(k_z,\mathbf{r}) 
\end{eqnarray}
where $\mathbf{e_z}$ is the so-called pilot vector, the unit vector pointing in the $z$ direction. These WVFs $\mathbf{M_n}(k_z, \mathbf{r})$, $\mathbf{N_n}(k_z, \mathbf{r})$ correspond to TE/TM modes of the field.

One can show  \cite{Chew1999} that the homogeneous part of the Green's function can be expanded in terms of these vector wavefunction in the following way:
\begin{eqnarray}
\mathbf{G}_{0}(\mathbf{r},\mathbf{r^\prime}, \omega) &=& -\dfrac{\mathbf{e_{\rho}e_{\rho}}}{k_0^2} \delta (\mathbf{r}-\mathbf{r^\prime}) 
\\
&+&
\dfrac{i}{8\pi}\sum_{n= - \infty}^{\infty} \int\limits_{-\infty}^{\infty} \frac{dk_z}{k_{0\rho}^2} \mathbf{F}_n (k_z, \mathbf{r}, \mathbf{r^\prime})
\nonumber
\end{eqnarray}
and the $\mathbf{F}_n (k_z, \mathbf{r}, \mathbf{s})$ function is given by
\begin{equation}
\begin{cases}
\mathbf{M}_n^{(1)} (k_z, \mathbf{r})\overline{\mathbf{M}}_n (k_z, \mathbf{r^\prime}) + \mathbf{N}_n^{(1)} (k_z, \mathbf{r})\overline{\mathbf{N}}_n (k_z, \mathbf{r^\prime})&  
\\
\mathbf{M}_n (k_z, \mathbf{r})\overline{\mathbf{M}}_n^{(1)} (k_z, \mathbf{r^\prime}) + \mathbf{N}_n (k_z, \mathbf{r})\overline{\mathbf{N}}_n^{(1)} (k_z, \mathbf{r^\prime})& 
\end{cases}
\end{equation}
here the first line holds for $\rho_r > \rho_{r^\prime}$ while the second one for $\rho_r < \rho_{r^\prime}$, and $k_0 = {\omega}/{c}$, $k_{0\rho } = \sqrt{k_0^2 - k_z^2}$ and the superscript $(1)$ in vector wave functions denotes that the Bessel function of the first kind $J_n(k_\rho \rho)$ should be replaced with the Hankel function of the first kind $H^{(1)}_n(k_\rho \rho)$. Here we provide the explicit form of WVF:
\begin{eqnarray}
\mathbf{M}_n(k_z,\mathbf{r}) &=& 
\begin{pmatrix}
\frac{in}{\rho} J_n(k_{0\rho } \rho)\\
- k_{0\rho } (J_n(k_{0\rho } \rho))'\\
0
\end{pmatrix} e^{i n \theta + i k_z z},
\nonumber\\ 
\mathbf{N}_n(k_z,\mathbf{r}) &=& 
\begin{pmatrix}
\frac{ik_z k_{0\rho}}{k} (J_n(k_{0\rho} \rho))'\\
-\frac{n k_z}{\rho k} J_n (k_{0\rho } \rho)\\
\frac {k_{0\rho}^2}{k} J_n (k_{0\rho} \rho)
\end{pmatrix} e^{i n \theta + i k_z z} 
\nonumber\\
\overline{\mathbf{M}}_n(k_z,\mathbf{r^\prime}) &=& 
\begin{pmatrix}
-\frac{in}{\rho^\prime} J_n(k_{0\rho } \rho^\prime)\\
- k_{0\rho } (J_n(k_{0\rho } \rho^\prime))'\\
0
\end{pmatrix}^T e^{- i n \theta^\prime - i k_z z^\prime},
\nonumber\\ 
\overline{\mathbf{N}}_n(k_z,\mathbf{r^\prime}) &=& 
\begin{pmatrix}
-\frac{ik_z k_{0\rho }}{k} (J_n(k_{0\rho } \rho^\prime))'\\
-\frac{n k_z}{\rho^\prime k} J_n (k_{0\rho } \rho^\prime)\\
\frac {k_{0\rho }^2}{k} J_n (k_{0\rho } \rho^\prime)
\end{pmatrix}^T e^{- i n \theta^\prime - i k_z z^\prime}\raisetag{10\baselineskip} \nonumber\\
{} 
\end{eqnarray}
where $J_n(k_\rho \rho)'$ corresponds to derivative with respect to the dimensionless argument.

Now having the integral representation of the homogeneous term of the Green's function, we can construct the scattering term in a similar fashion. Let us denote the medium outside the dielectric cylinder as $1$ and the medium inside as $2$. The particular form of the Green's tensor depends on the position of a source point $\mathbf{r^\prime }$: whether it is inside or outside the cylinder. As soon as we are interested in a situation when both source and receiver are outside the cylinder and in the latter we consider only the second case. Thus, the total Green's tensor can  be written as:
\begin{eqnarray}
\begin{cases}
\mathbf{G}^{11}(\mathbf{r},\mathbf{r^\prime},\omega) = \mathbf{G}^{11}_0(\mathbf{r},\mathbf{r^\prime},\omega) + \mathbf{G}^{11}_s(\mathbf{r},\mathbf{r^\prime},\omega), \\
\mathbf{G}^{21}(\mathbf{r},\mathbf{r^\prime},\omega) = \mathbf{G}^{21}_s(\mathbf{r},\mathbf{r^\prime},\omega), 
\end{cases}
\end{eqnarray}
here the two superscripts denote position of the receiver and the source point respectively and the two scattering parts of the Green's tensor has the following form:
\begin{eqnarray}
\mathbf{G}_{s}^{11}(\mathbf{r,r^\prime,\omega}) &=&
\frac{i}{8 \pi} \sum_{n= - \infty}^{\infty} \int\limits_{-\infty}^{\infty} \frac{dk_z}{k_{\rho1}^2} \mathbf{F}^{11 (1)}_{\mathbf{M};n, 1}(k_z,\mathbf{r})\overline{\mathbf{M}}_{n,1}^{(1)}(k_z,\mathbf{r^\prime}) 
\nonumber \\  
&+&\mathbf{F}^{11 (1)}_{\mathbf{N};n,1}(k_z,\mathbf{r})\overline{\mathbf{N}}_{n,1}^{(1)}(k_z,\mathbf{r^\prime}) ,
\nonumber\\
\mathbf{F}^{11 (1)}_{\mathbf{M};n,1}(k_z,\mathbf{r}) &=& R^{11}_{MM}\mathbf{M} ^{(1)}_{n,1}( k_z, \mathbf{r})+R^{11}_{NM} \mathbf{N}_{n,1}^{(1)} (k_z,\mathbf{r}) ,
\nonumber\\
\mathbf{F}^{11 (1)}_{\mathbf{N};n,1}(k_z,\mathbf{r}) &=& R^{11}_{MN} \mathbf{M} ^{(1)}_{n,1}(k_z, \mathbf{r})+R^{11}_{NN} \mathbf{N}_{n,1}^{(1)} (k_z,\mathbf{r}).
\end{eqnarray}
\begin{eqnarray}
\mathbf{G}_{s}^{21}(\mathbf{r,r^\prime,\omega}) &=&
\frac{i}{8 \pi} \sum_{n= - \infty}^{\infty} \int\limits_{-\infty}^{\infty}  \frac{dk_z}{k_{\rho1}^2} \mathbf{F}^{21}_{\mathbf{M};n,2}(k_z,\mathbf{r})\overline{\mathbf{M}}_{n,1}^{(1)}(k_z,\mathbf{r^\prime}) 
\nonumber \\ 
&+& \mathbf{F}^{21}_{\mathbf{N};n,1}(k_z,\mathbf{r})\overline{\mathbf{N}}_{n,1}^{(1)}(k_z,\mathbf{r^\prime}) ,
\nonumber\\
\mathbf{F}^{21}_{\mathbf{M};n,2}(k_z,\mathbf{r}) &=& R^{21}_{MM} \mathbf{M}_{n,2}(k_z, \mathbf{r})+R^{21}_{NM} \mathbf{N}_{n,2} (k_z,\mathbf{r}) ,
\nonumber\\
\mathbf{F}^{21}_{\mathbf{N};n,2}(k_z,\mathbf{r}) &=& R^{21}_{MN} \mathbf{M}_{n,2}(k_z, \mathbf{r})+R^{21}_{NN} \mathbf{N}_{n,2} (k_z,\mathbf{r}),
\end{eqnarray}
here the scattering Fresnel coefficients $R_{AB}^{ij}$ are introduced and the second subscript in the VWFs denotes that $k$ and $k_\rho$ should be replaced with their values inside the corresponding media $k_i = \veps_i(\mathbf{r},\omega)k_0$, $k_{\rho i} = \sqrt[]{k_i^2 - k_z^2}$. We should notice that unlike the case of the homogeneous term, here we have products of $\mathbf{M}$ and $\mathbf{N}$, which is due to the fact that the normal modes in our case have hybrid nature.

The form of the Fresnel coefficients mentioned above can be found by imposing the boundary conditions on the Green's tensor at the surface of the cylinder:
\begin{eqnarray}
{\begin{cases}
	\mathbf{e}_{\rho} \times [\mathbf{G}^{11}(\mathbf{r},\mathbf{r^\prime},\omega) - \mathbf{G}^{21}(\mathbf{r},\mathbf{r^\prime},\omega) ]|_{\rho_r = \rho_c} = 0, 
	\\
	\mathbf{e}_{\rho} \times \mathbf{ \nabla_r } \times [\mathbf{G}^{11}(\mathbf{r},\mathbf{r^\prime},\omega) - \mathbf{G}^{21}(\mathbf{r},\mathbf{r^\prime},\omega)]|_{\rho_r = \rho_c} = 0.
	\end{cases}}
	\nonumber\\
\end{eqnarray}
Solving for this, we can find the Fresnel coefficients $R_{AB}^{ij}$ and, finally, construct the scattering part of the Green's tensor $\mathbf{G}_s(\mathbf{r},\mathbf{r}^\prime,\omega)$. We provide the explicit expressions for the Fresnel coefficients below:
\begin{widetext} 
	\begin{eqnarray}
	 DT(k_z) &=& -\left( \dfrac{1}{k_{\rho2}^2} - \dfrac{1}{k_{\rho1}^2} \right)^2 k_z^2 n^2 + \left( \dfrac{(J_n(k_{\rho2}\rho_c))^\prime}{k_{\rho2}J_n(k_{\rho2}\rho_c)} - \dfrac{(H^{(1)}_n(k_{\rho1}\rho_c))^\prime}{k_{\rho1}H^{(1)}_n(k_{\rho1}\rho_c)}\right) \times 
	 \left( \dfrac{(J_n(k_{\rho2}\rho_c))^\prime k_2^2}{k_{\rho 2}J_n(k_{\rho2}\rho_c)} - \dfrac{(H^{(1)}_n(k_{\rho1}\rho_c))^\prime k_1^2}{k_{\rho1}H_n^{(1)}(k_{\rho1}\rho_c)}\right) \rho_c^2 
	\nonumber\\
	 R_{MM}^{11}(k_z) &=& \dfrac{J_n(k_{\rho1}\rho_c)}{H^{(1)}_n(k_{\rho1}\rho_c)} \Bigg[ \left( \dfrac{1}{k_{\rho2}^2} - \dfrac{1}{k_{\rho1}^2} \right)^2 k_z^2 n^2 - \left( \dfrac{(J_n(k_{\rho2}\rho_c))^\prime}{k_{\rho2}J_n(k_{\rho2}\rho_c)} - \dfrac{(J_n(k_{\rho1}\rho_c))^\prime}{k_{\rho1}J_n(k_{\rho1}\rho_c)} \right) \times 
	 \nonumber\\
	&&\left( \dfrac{(J_n(k_{\rho2}\rho_c))^\prime k_2^2}{k_{\rho2}J_n(k_{\rho2}\rho_c)} - \dfrac{(H^{(1)}_n(k_{\rho1}\rho_c))^\prime k_1^2}{k_{\rho1}H^{(1)}_n(k_{\rho1}\rho_c)} \right) \rho_c^2 \Bigg] \dfrac{1}{DT(k_z)} 
	\nonumber\\
	R_{NM}^{11}(k_z) &=& \dfrac{J_n(k_{\rho1}\rho_c)}{H_n^{(1)}(k_{\rho1}\rho_c)}\dfrac{1}{k_{\rho1}} \left( \dfrac{1}{k_{\rho1}^2} - \dfrac{1}{k_{\rho2}^2}\right) \left( \dfrac{(J_n(k_{\rho1}\rho_c))^\prime}{J_n(k_{\rho1}\rho_c)} - \dfrac{(H^{(1)}_n(k_{\rho1}\rho_c))^\prime}{H_n^{(1)}(k_{\rho1}\rho_c)}\right) \dfrac{k_1 k_z n \rho_c}{DT(k_z)} 
	\nonumber\\
	R_{MN}^{11}(k_z) &=& R_{NM}^{11} 
	\nonumber\\
	R_{NN}^{11}(k_z) &=& \dfrac{J_n(k_{\rho1}\rho_c)}{H^{(1)}_n(k_{\rho1}\rho_c)} \Bigg[ \left( \dfrac{1}{k_{\rho2}^2} - \dfrac{1}{k_{\rho1}^2} \right)^2 k_z^2 n^2 - \left( \dfrac{(J_n(k_{\rho2}\rho_c))^\prime}{k_{\rho2}J_n(k_{\rho2}\rho_c)} - \dfrac{(H^{(1)}_n(k_{\rho1}\rho_c))^\prime}{k_{\rho1}H^{(1)}_n(k_{\rho1}\rho_c)} \right) \times 
	\nonumber\\
	&&\left( \dfrac{(J_n(k_{\rho_2}\rho_c))^\prime k_2^2}{k_{\rho2}J_n(k_{\rho2}\rho_c)} - \dfrac{(J_n(k_{\rho1}\rho_c))^\prime k_1^2}{k_{\rho1}J_n(k_{\rho1}\rho_c)}\right)\rho_c^2\Bigg]\dfrac{1}{DT(k_z)}
	\end{eqnarray}
\end{widetext}
In order to extract the fundamental guided mode contribution to the Green's tensor, one needs to take the common denominator of all of the Fresnel coefficients and expand it near the corresponding $\beta_{\text{HE}_{11}}$ value up to the first order: $DT(k_z) \approx  \dfrac{\partial DT(k_z)}{\partial k_z} \bigg|_{k_z = \beta_{\text{HE}_{11}}} \left(k_z - \beta_{\text{HE}_{11}} \right) + ...$. Then one needs to calculate the pole contribution to the integral by using the residue theorem and finding the value of $\beta_{\text{HE}_{11}}$ from the dispersion relation.

\section{}
\label{App B}

Starting from the Eq. \eqref{ST}, where $S_{ii} = 1 - i \hbar \gamma_{wg}^{(f)} \sum_j \dfrac{f_j^{(t)}}{\hbar \Delta - \lambda_j}$, we want to express the transmission spectra $t = |S_{ii}|^2$ in a convenient way. For this, let us consider different kinds of terms:

\begin{eqnarray}
	& |S_{ii}|^2 = 1 + \sum\limits_{j=1}^{N} \bigg( \left| \hbar \gamma_{wg}^{(f)} \dfrac{f_j^{(t)}}{\hbar \Delta - \lambda_j} \right|^2 + 2 \hbar \gamma_{wg}^{(f)} \text{Im} \:\dfrac{f_j^{(t)}}{\hbar \Delta - \lambda_j} + \nonumber\\ & (\hbar \gamma_{wg}^{(f)})^2 \sum\limits_{i=1,i\ne j}^{N} \dfrac{f_j^{(t)} (f_i^{(t)})^*}{\left(\hbar \Delta - \lambda_j\right) (\hbar \Delta - \lambda_i^*)}\bigg)\nonumber\\
	& {}
\end{eqnarray}
The second term can be simply written into the form, similar to Eq. \eqref{ImTExp}:
\begin{eqnarray}
	\text{Im} \: \dfrac{f_j^{'(t)} + if_j^{''(t)}}{\hbar \Delta - \lambda_j' - i\lambda_j''} = \dfrac{ f_j'^{(t)} \lambda_j'' + f_j^{''(t)}(\hbar \Delta - \lambda_j') }{(\hbar \Delta - \lambda_j')^2 + \lambda_j''^2}.
\end{eqnarray}

The last term contains cross-products of contributions from different eigenstates having different eigenvalues and we want to rewrite it in a similar way, which can be done through a sequence of the following transformations: 
\begin{widetext}
	\begin{eqnarray}
	&\sum\limits_{j = 1}^{N} \sum\limits_{i = 1, i \ne j}^{N} \dfrac{f^{(t)}_j (f_i^{(t)})^*}{(\hbar \Delta - \lambda_j)(\hbar \Delta - \lambda_i^*)} = \sum\limits_{j = 1}^{N} \sum\limits_{i = 1, i \ne j}^{N} \dfrac{f_j^{(t)} (f_i^{(t)})^*}{\lambda_j - \lambda_i^*} \left[ \dfrac{1}{\hbar \Delta - \lambda_j} - \dfrac{1}{\hbar \Delta - \lambda_i^*} \right] = 
	\nonumber\\
	&\sum\limits_{j = 1}^{N} \sum\limits_{i = 1, i \ne j}^{N} \dfrac{f_{j}^{(t)}  (f_{i}^{(t)})^*}{\lambda_j - \lambda_i^*} \left( \dfrac{1}{\hbar \Delta - \lambda_j} \right) - \sum\limits_{i = 1}^{N} \sum\limits_{j = 1, j \ne i}^{N} \dfrac{f_{j}^{(t)} (f_{i}^{(t)})^*}{\lambda_j - \lambda_i^*} \left( \dfrac{1}{\hbar \Delta - \lambda_i^*} \right)  = \left| i \leftrightarrow j \: \text{for the 2nd term} \right| =
	\nonumber\\
	& \sum\limits_{i \ne j} 2 \text{Re} \: \dfrac{f_j^{(t)} (f_i^{(t)})^*}{\lambda_j - \lambda_i^*} \dfrac{1}{\hbar \Delta - \lambda_j} = \sum\limits_{i \ne j} 2 \text{Re} \: \left[ \dfrac{f_j^{(t)} (f_i^{(t)})^*}{\lambda_j - \lambda_i^*} \left( \hbar \Delta - \lambda_j^* \right) \right] \dfrac{1}{(\hbar \Delta - \lambda_j')^2 + \lambda_j''^2} = 
	\nonumber\\
	&\sum\limits_{i \ne j} \left\{ 2\text{Re} \left[ \dfrac{f_j^{(t)} (f_i^{(t)})^*}{\lambda_j - \lambda_i^*} \right] \left( \hbar \Delta - \lambda_j' \right) - 2\text{Im} \left[  \dfrac{f_j^{(t)} (f_i^{(t)})^*}{\lambda_j - \lambda_i^*} \right] \lambda_j'' \right\} \dfrac{1}{(\hbar \Delta - \lambda_j')^2 + \lambda_j''^2}. 
	\nonumber\\
	& {}
	\end{eqnarray}
\end{widetext}

Finally, by combining all terms together we seek the Eq. \eqref{TExp}. Similarly to this, Eq. \eqref{RExp} for the reflection coefficient can be obtained.


\end{document}